\documentclass[usegraphicx,useAMS,usenatbib]{mn2e}
\usepackage{times}
\usepackage{amssymb}
\usepackage{amsmath}
\usepackage{graphicx}
\usepackage{euscript}
\usepackage{rotating}
\usepackage{epsfig}
\usepackage{url}

\def\mhz{{\rm\thinspace MHz}}
\def\ghz{{\rm\thinspace GHz}}

\def\ergpspcmsq{{\rm\thinspace erg~s^{-1}~cm^{-2}}}
\def\ergpspHz{{\rm\thinspace erg~s^{-1}~Hz^{-1}}}
\def\ergps{{\rm\thinspace erg~s^{-1}}}
\def\mpc{{\rm\thinspace Mpc}}
\def\kpc{{\rm\thinspace kpc}}

\def\erg{{\rm\thinspace erg}}
\def\kmpspmpc{\hbox{$\rm\thinspace km~s^{-1}~Mpc^{-1}$}}
\def\kev{{\rm\thinspace keV}}

\newcommand{\eg}{e.g.\thinspace}
\def\chandra{{\it Chandra}}

\begin{document}

\title{The radio properties of a complete, X-ray selected sample of nearby, massive elliptical galaxies}
\author[Dunn et al.]
{\parbox[]{6.in} {R.J.H.Dunn$^{1,2}$\thanks{E-mail:
      robert.dunn@ph.tum.de}\thanks{Alexander von Humboldt Fellow}, S.W.Allen$^{3,4}$, G.B.Taylor$^5$\thanks{Adjunct
      Astronomer at the National Radio Astronomy Observatory},
    K.F.Shurkin$^5$, G.Gentile$^{6,7}$ A.C.Fabian$^8$ and C.S.Reynolds$^9$\\ 
    \footnotesize
    $^1$ Excellence Cluster ``Universe'', Technische Universit\"at
    M\"unchen, Boltzmannstrasse 2, D-85748, Garching, Germany\\
    $^2$ School of Physics and Astronomy, Southampton, University of
    Southampton, SO17 1BJ\\
    $^3$ Kavli Institute for Particle Astrophysics and Cosmology at Stanford
University, 452 Lomita Mall, Stanford, CA 94305-4085, USA\\
    $^4$ SLAC National Accelerator Laboratory, 2575 Sand Hill Road, Menlo Park, CA
94025, USA \\
    $^5$ University of New Mexico, Department of Physics and
    Astronomy, Alberquerque, NM 87131, USA. \\
    $^6$ Institut d'Astronomie et d'Astrophysique, Universit\'e Libre
    de Bruxelles, CP 226, Boulevard du Triomphe, B-1050 Bruxelles,
    Belgium\\
    $^7$ Sterrenkundig Observatorium, Universiteit Gent, Krijgslaan
    281, B-9000 Gent, Belgium \\
    $^8$ Institute of Astronomy, Madingley Road, Cambridge, CB3 0HA\\
    $^9$ Department of Astronomy and the Maryland Astronomy Center for
    Theory and Computation, University of Maryland, College Park, MD
    20742, USA\\ 
  }}

\maketitle

\begin{abstract}
We investigate the radio properties of a complete sample of
nearby, massive, X-ray bright elliptical and S0 galaxies.  Our sample
contains 18 galaxies with ROSAT All-Sky Survey X-ray fluxes $F_{\rm
X,\,0.1-2.4\kev}>3\times 10^{-12}\ergpspcmsq$, within a distance of
$100\mpc$.  For these galaxies, we have complete (18/18) VLA radio
and Chandra X-ray coverage.  Nuclear
radio emission is detected from 17/18 of the galaxies. Ten of the
galaxies exhibit extended radio emission; of these ten, all but one
also exhibit clear evidence of interaction of the radio source with
the surrounding, X-ray emitting gas. Among the seven galaxies with
unresolved radio sources, one has clear, and one has small,
cavity-like features in the \chandra\ X-ray images; a third has a
disturbed X-ray morphology.  Using a radio luminosity limit equivalent
to $L_{1.4\ghz}>10^{23}\rm{W~Hz^{-1}}$ to
calculate the radio-loud fraction, we find that this misses the
majority of the radio detected galaxies in the sample.  We determine integrated radio-to-X-ray
flux ratios for the galaxies, $\mathcal{GR}_{\rm X}$, which are shown
to span a large range (factor of 100). We calculate the mass-weighted
cooling times within $1\kpc$, and find hints for an anticorrelation
with the radio luminosity.  We also calculate limits on
$k/f$, where $k$ is the ratio of the total particle energy to that of
relativistic electrons radiating in the range $10\mhz$--$10\ghz$ and
$f$ is the volume filling factor of the plasma in the cavity. The
$k/f$ distribution is also broad, reflecting previous results for
larger galaxy clusters.  Lowering the X-ray flux limit, at the expense
of less complete VLA and Chandra coverage, increases the size of our
sample to 42 galaxies. Nuclear radio activity is detected in at least
34/42 of this extended sample.
\end{abstract}


\section{Introduction}\label{sec:intro}

Super-massive black holes (SMBH) appear to be common in
massive galaxies (see \eg \citealp{Richstone98,Ferrarese00,Kormendy01,Ferrarese02}). A question of
significant interest is the level to which these SMBHs are active
i.e. their emitted power as a fraction of accretion rate or Eddington
luminosity \citep{Best05,Nagar05,Chiaberge05,Balmaverde06,Capetti06,
  Filho06,Gallo08}. Generally, power is released from active galactic 
nuclei (AGN) in either radiative or kinetic forms. Whereas the radiated
power is straightforward to measure, using observations across the
electromagnetic spectrum, determining the kinetic power of AGN is more
challenging, particularly for low-to-moderate mass galaxies and
SMBHs. For the highest mass galaxies, however, and in particular those
at the centers of clusters and groups, nature provides powerful,
additional tools to probe AGN and their impact on their environments.

Within massive elliptical galaxies, groups and clusters, embedded AGN
have been shown to interact strongly with the surrounding hot X-ray
emitting medium. This interaction causes disturbances seen as shocks,
ripples, and cavities (\eg \citealp{Fabian00, Fabian03b, Fabian06,
Forman05}). The cavities, which are filled with low-density,
radio-emitting, relativistic plasma, appear as X-ray surface
brightness depressions -- a consequence of the X-ray luminous,
thermal gas being displaced by the jets of the AGN. X-ray cavities,
filled with radio emitting plasma, are clearly associated with the
central AGN of nearby clusters such as Perseus (\eg.,
\citealp{Bohringer93,Fabian00, Fabian03b, Fabian06}), Virgo (\eg
\citealp{Churazov01, Forman05}), Centaurus (\eg \citealp{Taylor06a})
and Hydra A (\eg \citealp{McNamara00}).

Crucially, studies of cavities allow the total kinetic energy and
kinetic power injected by AGN into their surroundings to be estimated
(e.g. \citealp{Churazov02, Allen06, Birzan04,
Rafferty06, Dunn06, Dunn08}). This injection is energetically
sufficient to prevent catastrophic cooling of the X-ray emitting gas, providing a potential
solution to the puzzle of why most nearby, massive galaxies appear
`red and dead' \citep{Bower06, Croton06, Croton06e, Somerville08}.
For a review of the effect SMBHs have on their host galaxy's evolution
see \citet{Cattaneo09}.

\citet{Dunn06, Dunn08} showed that almost all galaxies at
the centers of nearby clusters for which the central cooling time of
the X-ray emitting gas is short (i.e. $t_{\rm cool}<<t_{\rm H}$), and
for which this gas would otherwise cool rapidly and form stars, host
powerful, radio-emitting AGN and X-ray cavities, this extended the
work of \citet{Burns90} who showed that 70 
per cent of dominant galaxies in rich clusters with cooling cores are
radio loud.  Using a larger sample of clusters and groups,
\citet{Sun09} argue the inverse, that all brightest cluster galaxies
with radio AGN have cool cores.  As radio-emitting AGN are seen in almost all clusters with
short cooling times, the duty cycle of these AGN must be high.

It is important to recognise, however, that the radio sources at the
centres of clusters are in extreme environments. To assess the impact
of such processes on the general population of massive galaxies, it is
essential to extend this work to the general population of massive
ellipticals. Recent studies have used the SDSS\footnote{Sloan Digital
Sky Survey} \citep{York00, Stoughton02} and the NVSS \footnote{NRAO
(National Radio Astronomy Observatory) VLA (Very Large Array) Sky
Survey} \citep{Condon88} to examine the fraction of galaxies with
radio-loud AGN.  \citet{Best05} combine these two surveys, and select
galaxies with redshifts between $0.03<z<0.1$.
They show that the fraction of galaxies with a radio-loud
AGN depends on the stellar mass of the galaxy; the maximum fraction is
between 30 and 40 per cent, significantly below that inferred for
nearby clusters. (Note that the `maximum' fraction is obtained for the
least restrictive constraint on the radio luminosity and the largest
galaxies in that study.)  Other investigations into the fraction of
actively accreting SMBHs in the nearby Universe have led to similar
results, though do not rule out the possibility of significantly
larger fractions for lower radio luminosity thresholds (see
e.g. \citealp{Nagar05,Chiaberge05,Balmaverde06,Capetti06, Filho06}).  

Another angle has been taken by \citet{Goulding09} who select very local
($D<15\mpc$) galaxies which are bolometrically luminous using
infra-red observations.  In this volume limited sample (94 per cent
complete) of the 64 galaxies, 17 host an AGN (27 per cent).

In this paper we use high quality X-ray and radio observations to
determine the fraction of active AGN in a complete sample of nearby,
massive elliptical galaxies, selected to have high optical and X-ray
fluxes and lie within a distance of 100Mpc. Our sample includes both
dominant galaxies at the centres of groups and clusters and field
ellipticals with their own halos of hot X-ray gas. We show that the
fraction of galaxies that are active, i.e. exhibit radio
emission and/or clear X-ray cavities associated with a central AGN 
is very high: $\gtrsim 90$ per cent.  We also investigate the
prevalence of cool cores within this sample.  

Our discussion will proceed as follows. We begin in Section
\ref{sec:selection} with a description of the sample and selection
criteria. In Section \ref{sec:radio_data}, we describe the radio data
selection and processing. In Section \ref{sec:xray}, we describe the
X-ray data reduction and preparation. In Section \ref{sec:individ_sources}, we
discuss individual sources and in Section \ref{sec:population}, we
conduct a population study.  Section \ref{sec:extended} extends some aspects of the analysis to a
larger sample of galaxies, at the expense of incomplete VLA and Chandra
coverage.  We conclude in Section \ref{sec:concl}.

Throughout this paper, we assume a standard flat $\Lambda$CDM cosmology 
with $H_0$=70 km s$^{-1}$ Mpc$^{-1}$ and $\Omega_{\rm m}=0.3$.

\section{Sample Selection}\label{sec:selection}

Our study is focused on a clearly defined and essentially complete
sample of nearby elliptical galaxies with measured optical, X-ray and
radio properties.  We start from the catalogue of \citet{Beuing99}.  The
northern part of this catalogue contains all E and E/S0 galaxies from
the study of \citet{Faber89} with $\delta\geq0^{\circ}$ and magnitudes
brighter than $B_{\rm T}=13.5$ mag. The southern part of the
\citet{Beuing99} catalogue is drawn from \citet{Bender89}; it contains
all E and E/S0 galaxies with $\delta<0^{\circ}$ and $B_{\rm T}<13.5$
mag from \citep{Faber89}, as well as galaxies from the ESO
Lauberts-Valentijn Catalogue \citep{Lauberts89} brighter than $B_{\rm
T}=13.5$ mag and with morphological type $T_{\rm old}$ or $T_{\rm
new}\leq -3$.  In addition, the catalogue includes galaxies with
$\delta\geq0^{\circ}$, $B_{\rm T}<13.7$ and numerical class $T\leq-3$
from the Third Reference Catalogue for Bright Galaxies
\citep{deVaucouleurs91} and the Tully Nearby Galaxies Catalogue
\citep{Tully88}\footnote{\citet{Beuing99} adopt a fainter $B_{\rm T}$
threshold for the \citet{deVaucouleurs91} and \citet{Tully88}
catalogues, since \citet{deVaucouleurs91} found hints of a zero-point
offset of $0.12-0.18$ mag.}$^,$\footnote{The numerical class constraint
  excludes at least one well known X-ray bright galaxy, NGC\,1275, as
  it is classed as a peculiar galaxy, $T=99$}.  

The sample of \citet{Beuing99} contains 530 early type galaxies;
313 in the southern and 217 in the northern parts.  The distribution
of galaxies of different types is as follows: 49 per cent ellipticals
($T=-5$), one of which is a compact elliptical ($T=-6$);
15 per cent cD-type galaxies ($T=-4$); 26 per cent E/S0a ($T=-3$);
7 per cent S0s ($T=-2$); and 3 per cent later types
(\citealp{deVaucouleurs91}).  From their magnitude-distribution
function, \citet{Beuing99} judge that their sample starts to become
incomplete at $B_{\rm T}\approx13.0$.  However they state that it is
still 90 per cent complete at $B_{\rm T}=13.5$, assuming a
spatially homogeneous distribution.
 
\citet{Beuing99} provide X-ray luminosities, or upper limits on the
X-ray luminosity, for 293 galaxies in their
sample, based on data from the ROSAT\footnote{R\"ontgen Satellit ({\it
ROSAT})} All-Sky Survey (RASS). They exclude
galaxies which are not at the centre of the bright X-ray emission, but
embedded in it, {\it unless} they clearly stand out against it.  The
upper limits obtained for some galaxies were classed as meaningless if they
resulted from short exposures, or if the galaxy has a distance $>43h_{70}^{-1} \mpc$.  Their final sample consists of all
normal galaxies showing individual emission, as well as galaxies at
the centres of clusters and groups, if the X-ray emission is centred
on the galaxy and reasonably symmetric around it.  
\citet{OSullivan01} updated these $L_{\rm X}$ values, also incorporating pointed, follow-up X-ray observations made
with the ROSAT Position Sensitive Proportional Counter (PSPC), where
available. For our sample selection, we use the X-ray luminosities of
\citet{OSullivan01}, converted to our
reference cosmology. 207 galaxies in the \citet{OSullivan01} study have X-ray
luminosity measurements as opposed to upper limits.

From the initial catalogue of optically bright galaxies with X-ray
luminosity measurements, we identify all galaxies with $0.1-2.4$keV
fluxes $>3\times 10^{-12}\ergpspcmsq$. We also impose an upper
distance limit of 100 Mpc. Together, these two criteria ensure that
all galaxies studied should be sufficiently X-ray bright and nearby to
allow the properties of their diffuse, X-ray emitting gas to be
studied in detail with Chandra.

We note that the X-ray luminosities reported by \citet{OSullivan01}
are total luminosities, including the contributions from X-ray 
binaries, AGN and other compact sources, as well as the hot, gaseous
components. For the largest elliptical galaxies the
contributions from binaries is expected to be minimal (see
e.g. \citealp{Kim04, Humphrey08,Brough08}).   As we show in this
paper, the impact of the central AGN on the total X-ray emission is
also typically low.  Only in $\sim 10$ per cent (2/18) of our galaxies
does the AGN outshine the diffuse gas at X-ray wavelengths (see
Sections \ref{sec:individ_sources} and
\ref{sec:pop:RAct})\footnote{Two of the eighteen sources, IC\,310 and
  NGC\,4203, are anomalous in that their X-ray emission is dominated
  by their central AGN.  These two galaxies are also the two most
  late-type (S0) galaxies in the sample (see Table
  \ref{tab:sample}).  For these reasons, although IC\,310 and
NGC\,4203 formally meet the optical, X-ray and distance selection
criteria, we flag or exclude them from certain population studies.}
For the other galaxies, the X-ray emission from the central AGN
typically accounts for less than 1 per cent of the
total.

\begin{figure}
\centering
\includegraphics[width=1.0 \columnwidth]{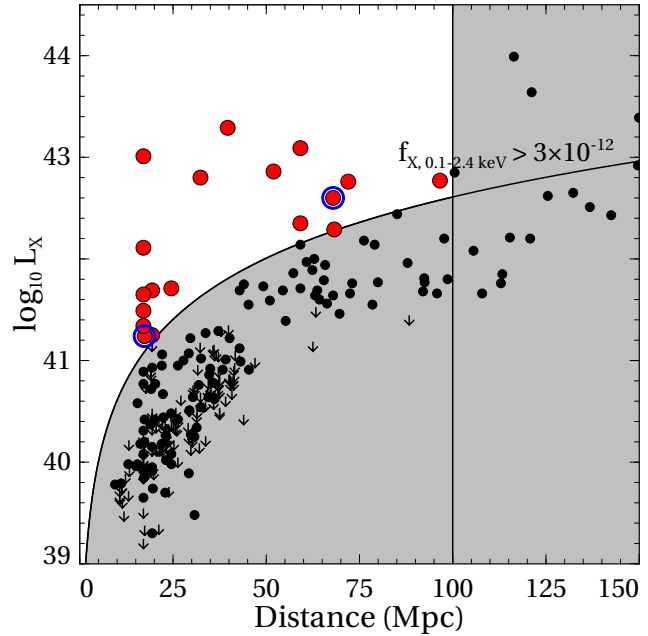}
\caption{\label{selectionfig} The distance-luminosity distribution of
  the galaxy sample of \citealt{Beuing99} 
with the flux and distance cuts used to obtain our final sample.  The 18
galaxies in our sample are highlighted in red. Galaxies for which only upper limits on
the X-ray luminosity are available are shown with arrows.  IC\,310
and NGC\,4203, for which the X-ray emission is dominated by the
central AGN, are highlighted with blue circles.  The X-ray luminosities
shown are taken from \citet{OSullivan01} but converted to
$H_0=70\kmpspmpc$.}
\end{figure}

The X-ray properties of the parent sample and X-ray flux and distance
cuts are shown in Fig. \ref{selectionfig}. The resulting sample
consists of 18 galaxies, which are detailed in Table \ref{tab:sample}.
VLA radio data and {\it Chandra} X-ray observations are available for
all 18 targets.

Two of the eighteen sources, IC\,310 and NGC\,4203, are anomalous in
that their X-ray emission is dominated by the central
AGN\footnote{The X-ray luminosities of \citet{OSullivan01}.
include the contributions of point sources in the galaxies.}
These two galaxies are the two most late type (S0) galaxies in the
sample (Table \ref{tab:sample}). For these reasons, although IC\,310
and NGC\,4203 formally
meet the optical, X-ray and distance selection criteria, we flag or
exclude them from certain population studies.

\begin{table*}
\caption{The Sample\label{tab:sample}}
\begin{tabular}{lccccrccc}
\hline
\hline
Source&Alternate
Name&RA$^a$&Dec$^a$&Redshift$^a$&Distance&\multicolumn{2}{c}{Type$^b$}&$\log(L_{\rm X})$\\
&&J2000&J2000&&Mpc&&&$\ergps$\\
\hline
IC\,310 & --- & 03h\,16m\,43.00s & +41d\,19m\,29.4s & 0.018940 &67.92& -2.0&-&42.60 \\
IC\,1860 & --- & 02h\,49m\,33.7s & -31d\,11m\,21s & 0.022902 &96.59& -4.7&BCG&42.77 \\
NGC\,499 & --- & 01h\,23m\,11.5s & +33d\,27m\,38s & 0.014673 &59.15& -2.8&-&42.35 \\
NGC\,507 & --- & 01h\,23m\,40.0s & +33d\,15m\,20s & 0.016458 &71.99& -3.2&-&42.76 \\
NGC\,533 & --- & 01h\,25m\,31.36s & +01d\,45m\,32.8s & 0.018509 &68.23& -4.8&-&42.29 \\
NGC\,708 & A262 & 01h\,52m\,46.48s &  +36d\,09m\,06.6s & 0.016195 &59.15& -4.8&BCG&43.09 \\
NGC\,1399 & --- & 03h\,38m\,29.08s & -35d\,27m\,02.7s & 0.004753 &19.40& -4.2&BCG&41.69 \\
NGC\,1404 & --- & 03h\,38m\,51.92s & -35d\,35m\,39.8s & 0.006494 &19.40& -4.7&-&41.25 \\
NGC\,1550 & --- & 04h\,19m\,37.93s & +02d\,24m\,35.7s & 0.012389 &51.95& -3.9&-&42.86\\
NGC\,4203 & --- & 12h\,15m\,05.06s & +33d\,11m\,50.38s & 0.003623 &17.38&-2.7&-&41.24 \\
NGC\,4406 & M\,86 & 12h\,26m\,11.74s & +12d\,56m\,46.40s & -0.000814&17.06&  -4.7&-&42.11\\
NGC\,4472 & M\,49 & 12h\,29m\,46.76s & +08d\,00m\,01.71s & 0.003326 &17.06& -4.7&-&41.49 \\
NGC\,4486 & M\,87 & 12h\,30m\,49.4s & +12d\,23m\,28s & 0.004360 &17.06& -4.3&BCG&43.01\\
NGC\,4636 & --- & 12h\,42m\,49.9s & +02d\,41m\,16s & 0.003129 &17.06& -4.8 &-&41.65\\
NGC\,4649 & M\,60 & 12h\,43m\,39.66s & +11d\,33m\,09.4s & 0.003726 &17.06& -4.6 &-&41.34\\
NGC\,4696& Centaurus Cluster & 12h\,48m\,49.3s & -41d\,18m\,40s &0.009867 &39.65& -3.9 &BCG&43.29\\
NGC\,5044 & --- & 13h\,15m\,23.97s & -16d\,23m\,07.9s & 0.009020 &32.36& -4.8&-&42.80 \\
NGC\,5846 & --- & 15h\,06m\,29.29s & +01d\,36m\,20.24s & 0.005717 &24.55& -4.7&-&41.71 \\

\hline
\end{tabular}
\begin{quote}
The sample of galaxies. $^a$ Positions
and redshifts obtained from NASA/IPAC Extragalactic Database (NED).
$^b$ The Type shows the $T$-type \citet{deVaucouleurs91} with
E=-5, E/S0=-3, S0=-2, S0a=0.   Also shown is whether the galaxy is a Brightest
Cluster Galaxy, (BCG), defined as the galaxy (likely) to be the dominant one in
a cluster listed in the Abell catalogues \citep{Abell58,Abell89}.  Also listed are the total X-ray luminosities 
(Section \ref{sec:selection}).
\end{quote}
\end{table*}

\section{Radio Data}\label{sec:radio_data}

The radio data for the targets were obtained  from the 
NRAO VLA archive.\footnote{The VLA (Very Large Array) is operated by the
National Radio Astronomy Observatory (NRAO).  The National Radio
Astronomy Observatory is operated by Associated Universities, Inc.,
under cooperative agreement with the National Science Foundation.}
Multiple observations of each source are available, for
various wavelengths and configurations.  For each source we 
select and present the radio map best suited to the present purposes.

In selecting the data to be analysed, preference was given to more
recent observations, and to those with time on source $\geq 5$
minutes.  As steep spectrum emission is brighter at lower frequencies,
preference was also given to observations performed at 1.5\ghz.  In addition, A-configuration observations at 1.5\ghz, or
B-configuration observations at 5~GHz were desirable in order to
provide arcsecond resolution for comparison with X-ray images from
\textit{Chandra}.

The data were reduced in the standard manner using {\scshape
aips}\footnote{Astronomical Image Processing System}\citep{Greisen03}. After an initial
editing of the data, absolute amplitude and phase calibration were
performed on each dataset using the scripts {\scshape vlaprocs} and
{\scshape vlarun}. For datasets in which the flux calibrator was
resolved, a model was used (if available) for the calibration. If bad
data were still present after the initial calibration, those data were
flagged and the calibration was repeated.

An image of each source was created using the task {\scshape
imagr}. Side-lobes from outlying sources were removed by using
multiple facets while imaging. Proper placement of the facets was
determined using the task {\scshape setfc}, which was set up to search
a 0.5 degree radius for sources in the NVSS catalogue with flux $\geq
10$ mJy. Typically, sources outside of that range are either not
bright enough or too far from the pointing centre to have an
appreciable effect on the quality of the image. In any event in which
{\scshape setfc} failed to include significant sources, facet
positions were set manually. For those sources with sufficient
signal-to-noise ratios, imaging and phase-only self-calibration were
then performed iteratively, until the theoretical noise was reached or
until the quality of the map ceased to benefit from the iterations.
Observational parameters and data for the VLA observations are
tabulated in Table \ref{tab:vla}.

For NGC\,1404 the VLA radio map from 2007 shows a faint point source
coincident with the galaxy centre.  However the detection is only just
above the level of the noise in the data, and so cannot be determined
accurately.  The NVSS catalogue also lists a source at the location of
NGC\,1404. We list this detection separately in Table \ref{tab:vla}.
During the analysis of the radio data for NGC\,4472 we found that the data from the A
configuration has a $(u,v)$ coverage such that the large scale emission
was not detected.  As a result of this, we checked all other
  A-configuration observations for undetected large scale emission
  using C-configuration data present in the VLA Archive.  This problem
was not found in any other galaxy apart from NGC\,4472.

The reduction and analysis of the $1.5\ghz$ radio data for NGC\,4486
are described by \citet{Hines89}.

\begin{table*}
\caption{VLA Data and Observational Parameters\label{tab:vla}}
\begin{tabular}{lccccccccr}
\hline
\hline
Source&Frequency&Config&Date&Time on Source&Peak Flux&Total Flux Density&RMS&Beam&PA \\
&(\ghz)&&&(sec)&(Jy/beam)&(Jy)&($\mu$Jy/beam)&(arcsec)&(deg)\\
\hline
IC\,310 & 4.885 & C & 1984 May 01 & 1170 & 1.36$\times10^{-1}$ & 1.98$\times10^{-1}$ & 73.9 & 3.99$\times$3.93 & -43.55 \\
IC\,1860 & 1.365 & A & 2007 Jun 08 & 8070 & 1.23$\times10^{-2}$ & 1.39$\times10^{-2}$ & 60.2 & 2.92$\times$1.42 & 11.83 \\
NGC\,499 & 1.365 & A & 2007 Jun 08 & 6070 & $<$1.18$\times10^{-4}$ & - & 35.8 & 1.43$\times$1.16 & -62.60 \\
NGC\,507 & 1.525 & C & 1984 May 14 & 1770 & 8.61$\times10^{-3}$ & 8.45$\times10^{-2}$ & 87.1 & 17.58$\times$12.11 & -81.43\\
NGC\,533 & 4.885 & C & 1997 Jul 05 & 230 & 5.06$\times10^{-3}$ & 8.91$\times10^{-3}$ & 93.3 & 5.08$\times$4.60 & -14.47 \\
NGC\,708 & 1.365  & A & 2000 Dec 13 & 1520 & 6.57$\times10^{-3}$ & 8.25$\times10^{-2}$ & 55.5 & 1.46$\times$1.18 & -49.36 \\
NGC\,1399 & 1.465 & H$^1$ & 1983 Dec 16 & 27910 & 1.89$\times10^{-2}$ & 4.63$\times10^{-1}$ & 72.1 & 4.15$\times$2.80 & 38.46 \\
NGC\,1404 & 1.365 & A & 2007 Jun 08 & 6060 & 3.50$\times10^{-4}$ & - & 85.8 & 3.64$\times$1.48 & 10.82 \\
NGC\,1550 & 1.365 & C & 1997 Sep 18 & 410 & 2.68$\times10^{-3}$ & 1.27$\times10^{-2}$ & 76.7 & 18.6$\times$16.8 & -84.14 \\
NGC\,4203 & 4.885 & A & 1999 Sep 05 & 423 & 1.66$\times10^{-2}$ & 1.80$\times10^{-2}$ & 93.6 & 0.48$\times$0.39 & 63.12 \\
NGC\,4406 & 4.885 & C & 1984 Jun 03 & 510 & 5.09$\times10^{-4}$ & 4.91$\times10^{-4}$ & 67.1 & 4.32$\times$3.86 & 55.55 \\
%NGC\,4472 & 1.425 & A & 2002 Mar 12 & 7500 & 4.65$\times10^{-2}$ & 1.29$\times10^{-1}$ & 68.1 & 4.04$\times$1.49 & -31.90\\
NGC\,4472 & 1.489 & C & 1986 Nov 26 & 192 & 1.19$\times10^{-1}$ & 2.23$\times10^{-1}$ & 250 & 17.19$\times$14.24 & -37.6 \\
NGC\,4486 & 1.466 & A & 1989 Mar 01 & 10020 & 1.74$\times10^{\; 0}$ & 1.38$\times10^{\; 1}$ & 32.5 & 0.80$\times$0.80 & 0.00\\
NGC\,4636 & 1.425 & A & 2002 Mar 12 & 12440 & 4.80$\times10^{-3}$ & 6.45$\times10^{-2}$ & 45.8 & 1.42$\times$1.33 & -18.42\\
NGC\,4649 & 1.465 & B & 1984 Jan 24 & 4410 & 1.72$\times10^{-2}$ & 2.82$\times10^{-2}$ & 29.7 & 4.51$\times$3.64 & 44.76 \\
NGC\,4696 & 1.565 & A & 1998 Apr 23 & 3635 & 3.58$\times10^{-1}$ & 3.64$\times10^{\; 0}$ & 118 & 4.40$\times$2.09 & 0.50\\
NGC\,5044 & 1.465 & A & 1992 Nov 27 & 8560 & 2.71$\times10^{-2}$ & 2.99$\times10^{-2}$ & 32.8 & 1.77$\times$1.03 & -16.83 \\
NGC\,5846 & 1.465 & A & 2002 Mar 12 & 9220 & 9.93$\times10^{-3}$ & 1.02$\times10^{-2}$ & 20.8 & 1.32$\times$1.18 & -17.86 \\
\hline
NGC\,1404$^\dagger$ & 1.400 & D & 1993 Oct 07 & - &3.42$\times10^{-4}$ & 3.90$\times10^{-4}$  & 450 & 45$\times$45 & 0 \\
\hline
\end{tabular}
\begin{quote}
$^1$ Hybrid configuration (A and B). $^\dagger$ From NVSS,as the point
  source in the A-configuration image is only just above the noise.
\end{quote}
\end{table*}

\section{X-ray Data Reduction}\label{sec:xray}

\begin{table*}
\caption{X-ray data parameters\label{tab:xray}}
\begin{tabular}{lcccrccl}
\hline
\hline
Source&Date&Obs ID&Detector&Mode&Exposure$^1$&$N_{\rm H}^\dagger$\\
\hline
IC\,310  & 23 Dec 2004& 5597 & ACIS-I& FAINT&  25.2 & 1.30 \\
IC\,1860 & 12 Sep 2009 & 10537 & ACIS-S&VFAINT &  37.2 & 2.40 \\
NGC\,499 & 4 Feb 2009 & 10865 & ACIS-S& VFAINT&  5.1 & 5.30 \\
 & 5 Feb 2009 & 10866 & ACIS-S& VFAINT&  8.0 & 5.30 \\
 & 7 Feb 2009 & 10867 & ACIS-S& VFAINT&  7.0 & 5.30 \\
 & 12 Feb 2009 & 10536 & ACIS-S& VFAINT&  18.4 & 5.30 \\
NGC\,507 & 11 Oct 2000& 2882 & ACIS-I& VFAINT& 43.6 & 5.24 \\
NGC\,533 & 28 Jul 2002& 2880 & ACIS-S& VFAINT&  34.8 & 3.10 \\
NGC\,708 & 03 Aug 2001& 7921 & ACIS-S& VFAINT& 110.6& 5.37 \\
NGC\,1399 & 18 Jan 2000& 319 & ACIS-S& FAINT& 55.9 & 1.34 \\
NGC\,1404 & 13 Feb 2003& 2492 & ACIS-S& FAINT&  29.0 & 1.45 \\
NGC\,1550 & 08 Jan 2002& 5800 & ACIS-S& VFAINT& 44.5 & 11.50 \\
NGC\,4203 & 10 Mar 2009& 10535 & ACIS-S& VFAINT& 41.4 & 1.10 \\
NGC\,4406 & 07 Apr 2000& 318  & ACIS-S& FAINT& 12.6 & 2.62 \\
NGC\,4472 & 12 Jun 2000& 321  & ACIS-S& VFAINT& 29.6 & 1.66 \\
NGC\,4486 & 05 Jul 2002& 2707  & ACIS-S& FAINT& 98.4 & 2.54 \\
NGC\,4636 & 26 Jan 2000& 323  & ACIS-S& FAINT& 44.2 & 1.81 \\
NGC\,4649  & 30 Jan 2007& 8182  & ACIS-S&  VFAINT& 49.2 & 2.20\\
 & 20 Apr 2000& 785  & ACIS-S&  VFAINT& 19.4 & 2.20\\
NGC\,4696$^2$& 01 Apr 2004& 4954 & ACIS-S& FAINT& 88.0 & 8.06 \\
NGC\,5044 & 19 Mar 2000& 3225 & ACIS-S&  VFAINT& 83.1 & 4.93 \\
NGC\,5846 & 24 May 2000& 788 & ACIS-S&  FAINT& 22.1 & 4.26 \\

\hline
\end{tabular}
\begin{quote}
$^1$ The final exposure in ks after all screening.  $^2$ Also analysed in
\citet{Taylor06a}. $^\dagger$ The Galactic column density, $N_{\rm
  H}$, is in units of $10^{20}\,atom{\rm
cm}^{-2}$ determined from H{\scshape i} studies \citep{Dickey90}.
\end{quote}
\end{table*}

The {\it Chandra} X-ray data were cleaned and reprocessed using the
{\scshape ciao} software and latest calibration files ({\scshape ciao}
v4.1.2, {\scshape caldb} v4.1.1). Hot pixels and cosmic ray afterglows
were identified and excluded. Charge-Transfer Inefficiency was
corrected for and standard grade selection applied.  Light curves were
generated from relatively source-free regions of the detectors and
used to screen the data for background flares. For observations made
in Very Faint (VF) mode, the additional information available for
identifying and screening background events was utilised.  Background
files were generated from the CALDB blank-field data-sets. These were
processed in a similar manner to the target observations.  The
background data-sets were normalised by the ratio of the $9-12\kev$
count rates in the source and background data-sets. Point-sources were
identified using the {\scshape wavdetect} wavelet-transform procedure,
and excluded.

To model other, possible background components not accounted for in
the blank-sky fields, background spectra were also extracted from relatively 
source-free regions of the detectors (e.g. the
ACIS-S1 chip for ACIS-S data) and compared with blank-sky fields.
Any excess soft emission components so detected were fitted with a
single temperature {\scshape mekal} model with a fixed solar
abundance and scaled appropriately in the subsequent analysis.
NGC\,5846 exhibits clear excess soft background emission.  Note,
however, that care is required here since in many cases the emission
from the cluster/group extends onto the S1 chip.  

For each galaxy, concentric annular regions were selected to give 
approximately constant signal-to-noise in the $0.6-7.0$ keV band.
Annuli were centered on the peak of the X-ray emission. Where the
Chandra data revealed the presence of a central point source, the
minimum radius was adapted to exclude it.  The signal-to-noise ratio for
the annuli was adjusted iteratively to obtain a number of 
regions ranging from 10 to 25.

The $0.6-7.0$ keV spectrum for each spherical shell was modelled as a
single temperature plasma, using the {\scshape mekal}
(e.g. \citealt{Mewe95}) model, with a {\scshape phabs} absorption
model fixed to the galactic $N_{\rm H}$.  The spectra for all annuli
were modelled simultaneously to determine the deprojected gas
temperature, metallicity and density profiles for the galaxies. The
modified C-statistic available in XSPEC \citep{Arnaud96,Arnaud04} was
used in all spectral fitting.

\begin{figure*}
\centering
\includegraphics[width=0.9\textwidth]{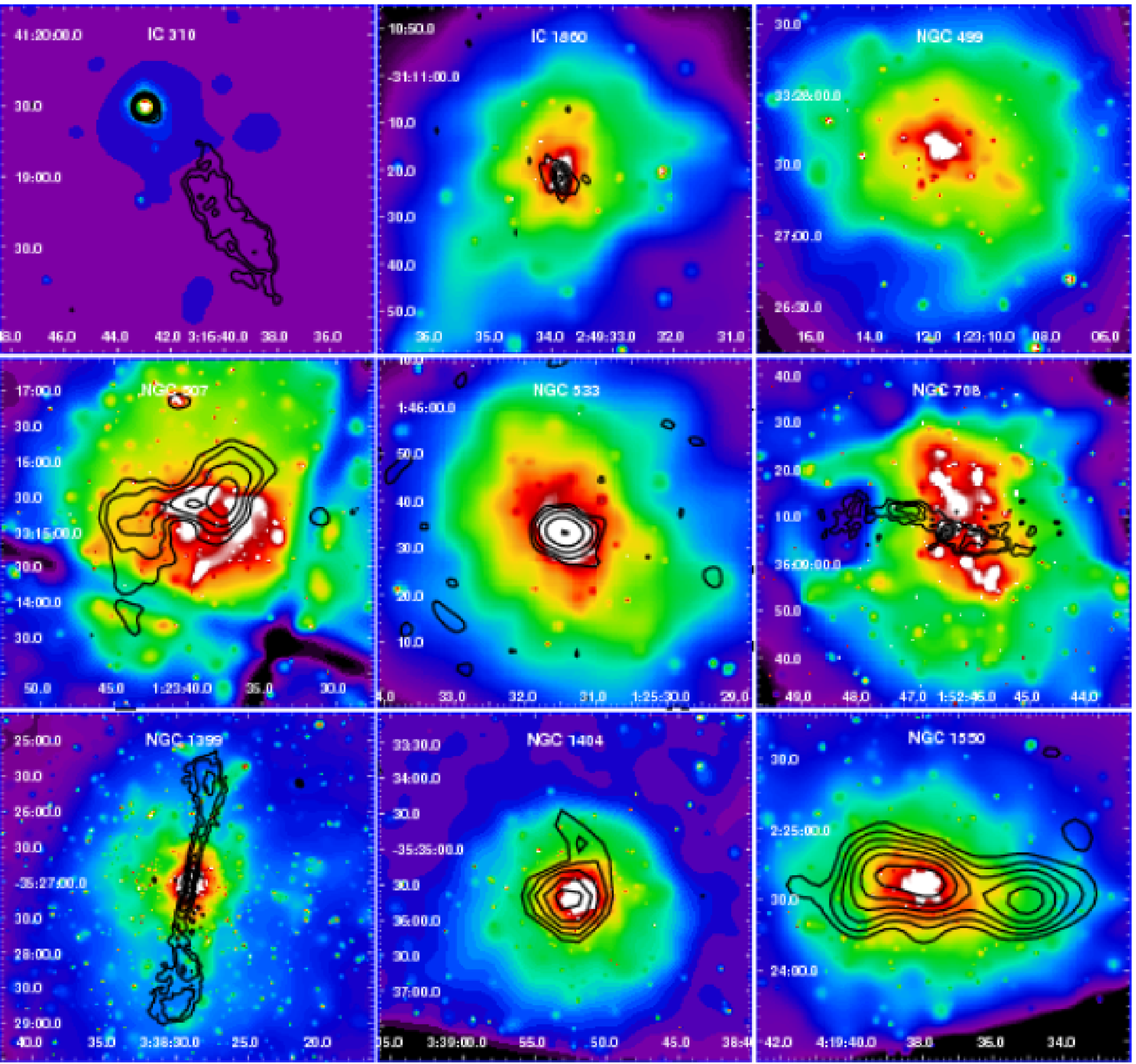}
\caption{Adaptively smoothed X-ray images (black/purple for faint
  emission, red/white for bright emission) with radio
  contours overlaid. From top left to
  bottom right: IC~310, IC~1860, NGC~499, NGC~507, NGC~533, NGC~708,
  NGC~1399, NGC~1404 and NGC~1550.  The Chandra image of NGC507 also
  shows artefacts associated 
  with chip gaps in the ACIS-I array.  The radio contours for
  NGC\,1404 are those from the NVSS survey.}
\label{fig:images}
\end{figure*}
\addtocounter{figure}{-1}
\begin{figure*}
\centering
\includegraphics[width=0.9\textwidth]{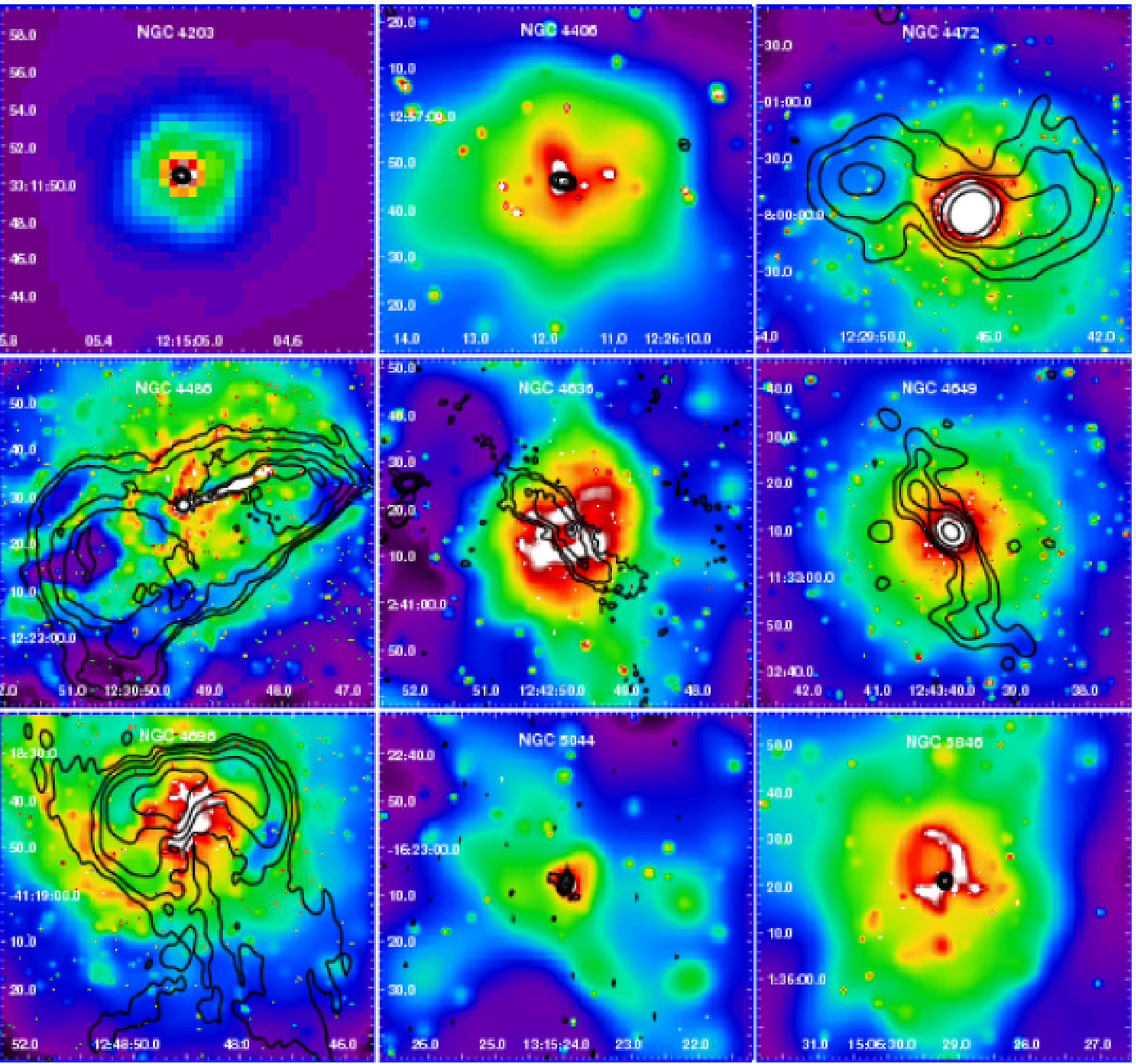}
\caption{(cont.)  From top left to
  bottom right: NGC~4203, NGC~4406, NGC~4472, NGC4486, NGC~4636,
  NGC~4649, NGC~4969, NGC~5044 and NGC~5846.}
\end{figure*}

\section{Notes on Individual Sources}\label{sec:individ_sources}

Our sample is essentially unbiased with respect to the morphology of
the X-ray emission from the galaxies.  X-ray emission from the hot,
diffuse gaseous halos and any AGN that may be present will both
contribute to the measured X-ray flux. In this Section, we discuss the
X-ray and radio morphologies for the galaxies in the sample, comment
on the presence of nuclear X-ray emission, and examine the evidence
for interactions between the central radio sources and surrounding
X-ray gas.  Several of the galaxies in the sample have previously been
studied by other authors.

\subsection{IC\,310}

IC\,310 is an S0 galaxy in the south-west region of the Perseus
cluster. The radio morphology is that of a head-tail galaxy
(e.g. \citealp{Sijbring98}) with a bright core and extended tail of
length around $400\kpc$. Our data show the radio tail disconnected
from the much brighter head.  There is only an observation at
$4.9\ghz$ which has sufficient resolution to show the structure of the source.

The X-ray emission is dominated by a central point source, coincident
with the radio core.  The extended X-ray halo is faint, reflecting the
relatively late-type nature of the galaxy. A small X-ray extension is
observed in the same direction as the radio tail. The current data do not
allow any detailed study of interactions between the radio and
X-ray emitting plasmas.  

\subsection{IC\,1860}

IC\,1860 is the dominant galaxy of the IC\,1860 group, which is itself
part of Abell S301.  The dominant galaxy of Abell S301 according to
\citet{Hudson01} is the Seyfert 2 galaxy IC\,1859, which is 6.5 arcminutes distant from
IC\,1860.  However, {\it XMM-Newton} data (Observation 0146510401)
shown in Fig. \ref{fig:images}, IC\,1860 sits at the
peak of the diffuse X-ray emission, and the position of IC\,1859 is
off to the west. 

\begin{figure}
\centering
\includegraphics[width=0.9\columnwidth]{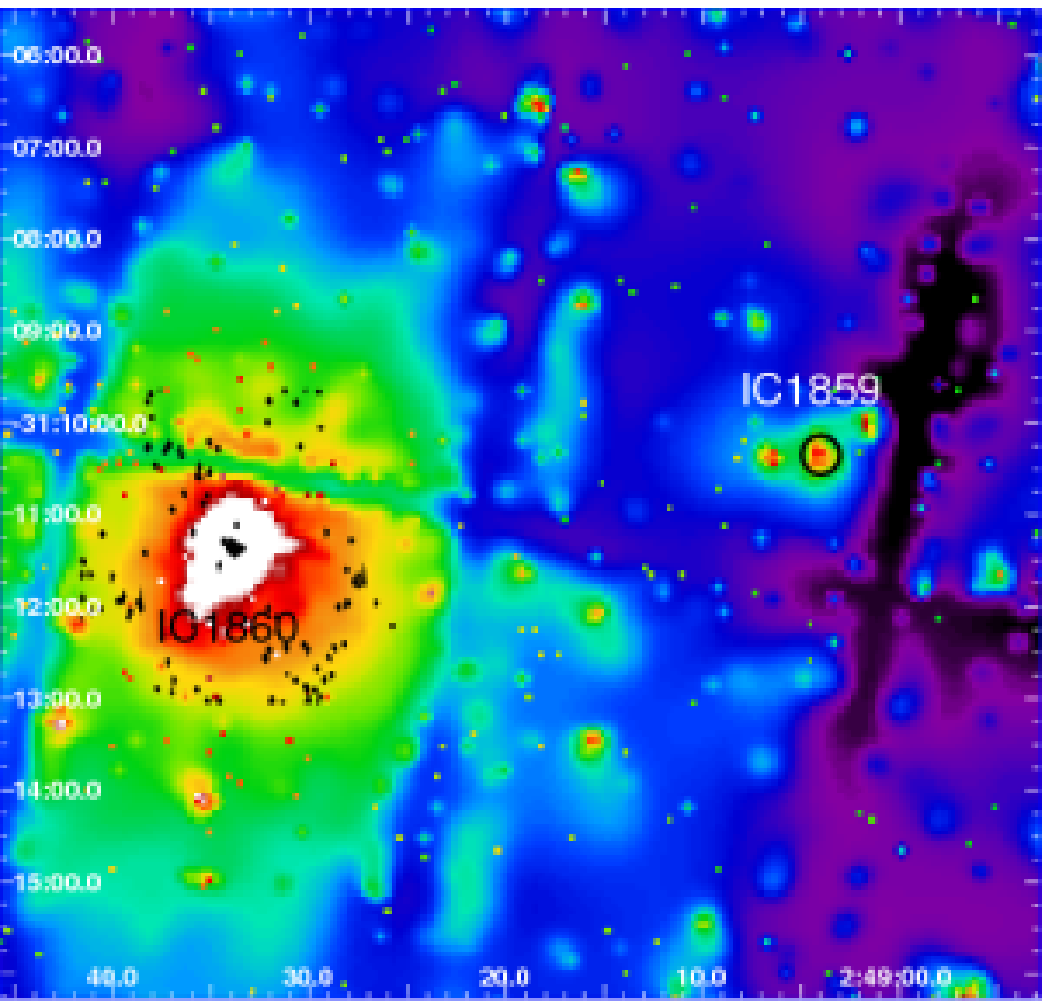}
\caption{The adaptively smoothed {\it XMM-Newton} X-ray emission
  surrounding IC\,1860 using the same
  colour scheme as Fig. \ref{fig:images},
  showing the location of IC\,1859 to the west.  The radio contours
  from the observation of IC\,1860 are overlaid.  The dominant galaxy,
  at least according to the X-ray emission, appears to be IC\,1860.
  }
\label{fig:ic1859}
\end{figure}

The radio data of IC\,1860 show a clear point source with faint extensions in the
northeast and southwest directions.  The X-ray data show an extension
towards the southeast.  However there is no clear correlation between
the faint radio extensions and the X-ray gas.

\subsection{NGC\,499}

NGC\,499 lies near NGC\,507 in the NGC\,507 galaxy Group.  There is no
radio detection for this galaxy down to a level of $1.18 \times
10^{-4}$ Jy, identifying NGC\,499 as abnormally radio quiet for its
X-ray luminosity.  The RMS noise in the VLA data is $35.8$ $\mu$Jy/beam,
which is fairly typical of the observations presented here (see Table
\ref{tab:vla}).  There appears to be a depression in the X-ray
emission to the south-east, visible in Fig. \ref{fig:images}.
However, whether this is the result of a past outburst of AGN activity  
is not clear.  \citet{Kim95, Paolillo03} report evidence for tidal
interactions between NGC\,499 and NGC\,507.

\subsection{NGC\,507}

NGC\,507 is the brightest member of a group of galaxies that is part
of the Pisces Cluster. Our radio map has comparatively poor spatial
resolution but show that the source is highly extended with two bright
lobes in a roughly east-west configuration.  {\it ROSAT} X-ray
observations showed NGC\,507 to have an extended, bright X-ray halo
with a luminosity comparable to that of poor clusters \citep{Kriss83,
Kim95}.  The {\it Chandra} data show some disturbance at the centre of
the cluster with indications of a shell of enhanced emission around
the eastern radio lobe \citep{Kraft04}. However, any signs of interaction are less
clear around the western lobe, though the secondary western, X-ray
bright lump may have been uplifted by the action of the AGN \citep{Forman01}.

\subsection{NGC\,533}

\begin{figure}
\centering
\includegraphics[width=0.9\columnwidth]{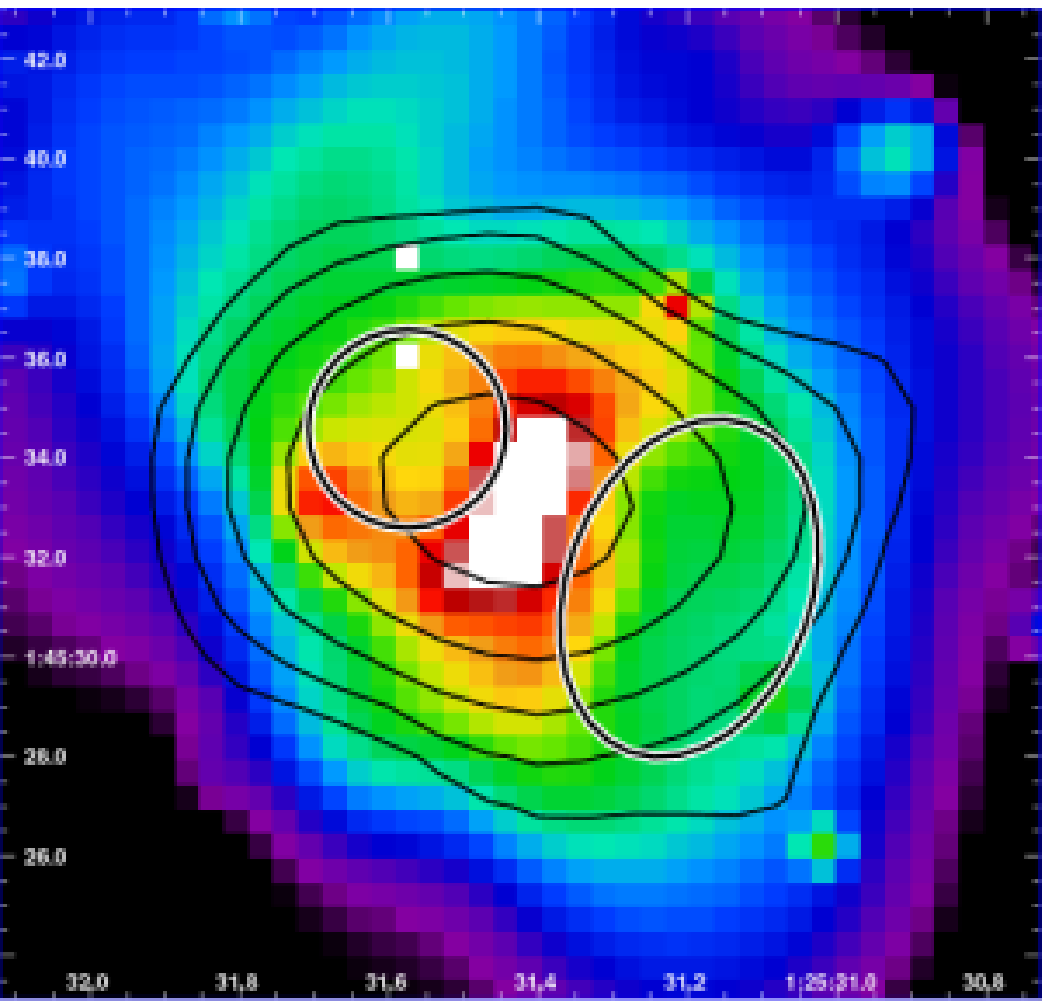}
\caption{Adaptively smoothed X-ray image of NGC\,533 using the same
  colour scheme as Fig. \ref{fig:images} with the radio
  contours overlaid in black, and the two depressions in the X-ray
  emission mentioned in the
  text highlighted.  Their radii/semi-major axes are $0.88$ for
  the north-eastern and
  $0.92\times 1.31\kpc$ for south-western cavities. }
\label{fig:ngc533}
\end{figure}

NGC\,533 is the dominant galaxy of a group of the same name.  Our
$4.9\ghz$ VLA
radio data show that the radio emission is dominated by a central
point source, although some extension to the west and southwest is
detected at modest significance.  The X-ray peak is coincident with
the central radio source (see \citealp{Sun09,Gastaldello07, Piffaretti05}).  On smaller scales, two crescent shaped
depressions in the X-ray emission are observed, either side of the
 peak, with a roughly northeast-southwest orientation (see also
 Fig. \ref{fig:ngc533}). The resolution of the current radio data is
 insufficient to 
determine unambiguously whether these features are the result of AGN
interaction. The cavities are therefore classified as `possible'
rather than `definite'. The larger scale X-ray emission also exhibits
northeast--southwest elongation, with a bay-like feature to the
southwest.  Low-frequency radio observations would be required to
investigate whether this bay is associated with a past generation of
AGN activity.

\subsection{NGC\,708}

NGC\,708 is the brightest galaxy of the Abell 262 cluster and hosts
the radio source B2 0149+35.  The VLA data reveal a central point
source and two lobes in an east-west configuration.  The X-ray
emission from the cluster is described by \citet{Blanton04}.  A
combined study of low frequency radio and X-ray data is presented by
\citet{Clarke09}, showing the interaction of the radio source with its
surroundings.

\subsection{NGC\,1399}

NGC\,1399 is the dominant galaxy in the Fornax Cluster (Abell S0373). The radio and
X-ray properties are described by \citet{Shurkin07}. The
radio source exhibits a nucleus with two 
tightly collimated jets that feed into more
extended radio lobes.  The X-ray
emission is peaked on the central AGN. Regions of enhanced X-ray
emission are observed just beyond the radio lobes \citep{Shurkin07}.

\subsection{NGC\,1404}

NGC\,1404 is a member of the Fornax Cluster, in the process of falling
towards the centre (NGC\,1399, \citealp{Machacek05}).  Our VLA
A-configuration data show a faint central point
source. The radio source is also clearly detected in the NVSS
catalogue. The X-ray emission map is fairly symmetric and also does
not show any features that would indicate significant interaction
between the central AGN and surrounding X-ray gas \citep{Machacek05}.

\subsection{NGC\,1550} \label{sec:ngc1550}

NGC\,1550, also known as NGC\,1551, is the dominant galaxy of the
NGC\,1550 group.  The source exhibits an unusual radio morphology with
two peaks: one approximately centered on NGC\,1550 and the other offset $\sim 45$
 arcsec to the west. The group is unusually X-ray luminous
for its temperature and velocity dispersion. The X-ray emission is
elliptical, with the semi major axis in the same direction as the axis
of the radio source \citep{Sun03}.  A filament of X-ray emission traces the eastern
radio lobe, wrapping around the southern part.  The western radio peak
appears uncorrelated with the X-ray emission. It is possible that the
western radio peak is actually due to a background source, although
the NASA\footnote{National Aeronautics and Space Administration}
Extragalactic Database\footnote{NASA/IPAC Extragalactic Database (NED)
is operated by the Jet Propulsion Laboratory, California Institute of
Technology, under contract with the National Aeronautics and Space
Administration.} lists no source at this position. It is also possible
that radio emission from NGC\,1550 has expanded in the surrounding ICM
in a highly asymmetric way to create the western peak.  

\subsection{NGC\,4203}

NGC\,4203 is an S0 galaxy hosting a low-luminosity low-ionisation
nuclear emission region (LINER) type AGN (LLAGN) \citep{Iyomoto98}.
Point like radio emission is detected in our $4.9\ghz$ VLA data.  The \chandra\
observation shows an X-ray structure similar to IC\,310.  There is
very little diffuse emission, with most of the X-rays coming from a
point source coincident with the radio emission.

\subsection{NGC\,4406}

Also known as M\,86, NGC\,4406 is the dominant member of a small group
falling at high velocity towards the centre of the Virgo cluster
(NGC\,4486/M\,87). Our $4.9\ghz$ VLA radio data reveal a central point source.
{\it XMM-Newton} and {\it Chandra} observations of the galaxy are
discussed by \citet{Finoguenov04} and \citet{Randall08}, respectively.
Our Chandra image shows only the core of the X-ray emitting gas.
There are no clear X-ray features indicating an interaction between the
radio and X-ray plasmas.

\subsection{NGC\,4472}

Also known as M\,49, this Seyfert 2 galaxy is the optically
brightest elliptical in the Virgo cluster.  It is also the dominant
member of a small group.  The central radio source
has a typical FRI structure but is unusually weak \citep{Ekers78}.  We
present the data from the VLA in C configuration.  As noted in Section
\ref{sec:radio_data}, the data from the A
configuration has a $(u,v)$ coverage such that the large scale emission
is not detected, and hence we present the C configuration data.  The VLA data
is a comparatively short observation, but the morphology matches that
seen in the FIRST\footnote{Faint Images of the Radio Sky at Twenty
  Centimetres, http://sundog.stsci.edu/ \citep{Becker95}} survey
observation, presented in 
\citet{Biller04, Allen06}.  There are cavities
in the X-ray emitting gas corresponding to the radio lobes \citep{Biller04}.

\subsection{NGC\,4486}

NGC\,4486, also known as M\,87 (Virgo A), is the dominant galaxy at the centre
of the Virgo Cluster. It hosts the well studied radio-bright AGN
3C\,274.  A bright radio jet is also clearly visible in the VLA
images, although the contours plotted have been chosen to highlight
the diffuse emission surrounding the core. The jet is also clearly
seen at X-ray wavelengths. The large scale structure of the ICM and
radio emission in NGC\,4486 is complex (see
e.g. \citealp{Hines89,Owen00,Young02,DiMatteo03,Forman05,Kovalev07,Forman07}).

\subsection{NGC\,4636}

NGC\,4636 (also known as NGC\,4624) lies in the southern regions of
the Virgo cluster, some 
$2.6\mpc$ from NGC\,4486. The radio data show a central core and two
clear lobes. The X-ray morphology is highly disturbed and suggests
that a much larger AGN outburst has occurred in the past (see also
\citet{Jones02, OSullivan05, Baldi09}).

\subsection{NGC\,4649}

\begin{figure}
\centering
\includegraphics[width=0.9\columnwidth]{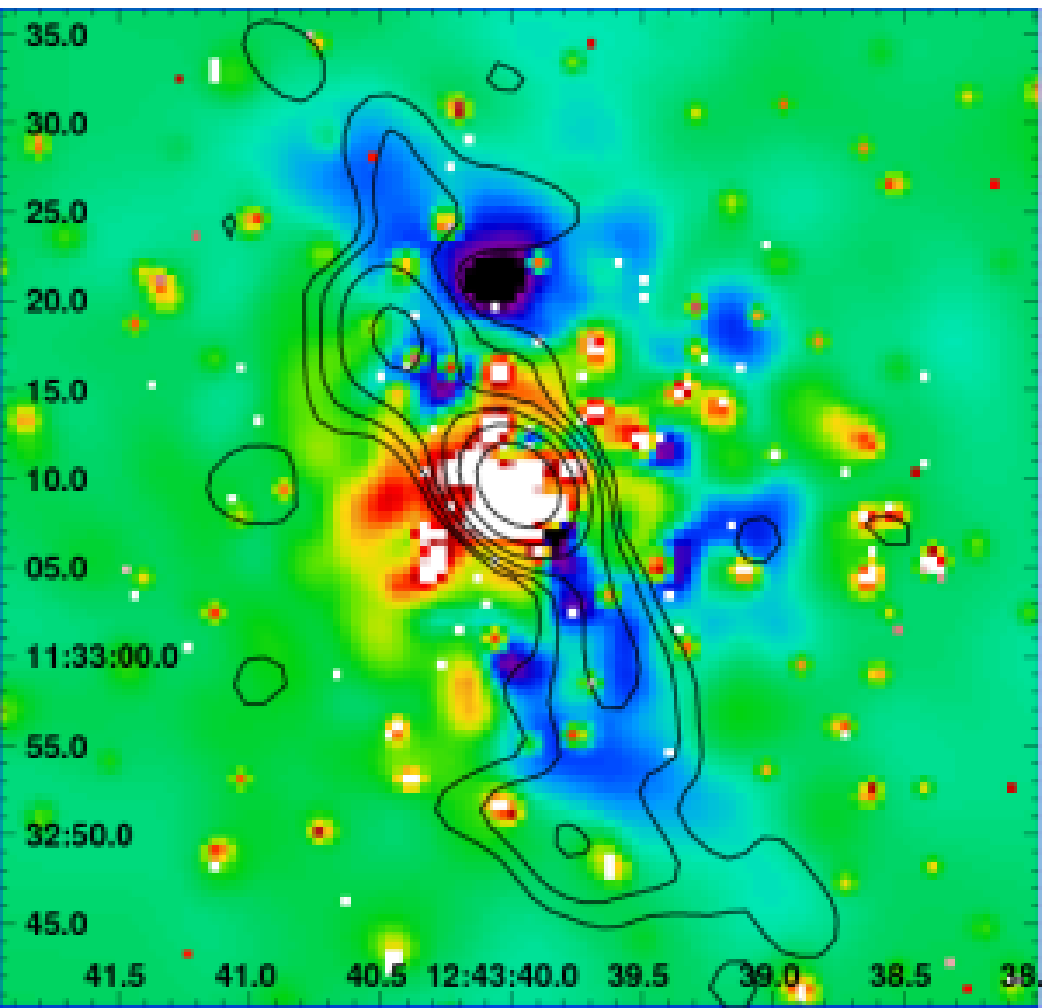}
\caption{The central regions of NGC\,4649 where a spherical (beta)
  model subtracted from the adaptively smoothed image.  The colour
  scheme is the same as in Fig. \ref{fig:images}.  The
  depressions associated with the radio emission (black contours) are clearly visible
  to the north and south of the core.}
\label{fig:ngc4649}
\end{figure}

NGC\,4649, also known as
M\,60 is located in a group at the eastern edge of the Virgo Cluster.
Recent detailed analyses of the radio and X-ray properties of this
source can be found in \citet{Randall04,Shurkin07,Humphrey08}.  In the deeper Chandra data
presented here, faint fingers of X-ray 
emission are observed to surround 
the radio emission, indicating a channel carved by the
AGN jet. Cavities are seen in the X-ray gas
at positions correlated with the
radio emission from the AGN, confirming the results of
\citet{Shurkin07}. These appear as ``wedges'' pointing to
the surface brightness peak in Fig. \ref{fig:images}.  To clarify the
location of the cavities we subtract a spherical profile (beta model)
which is fitted to the adaptively smoothed image.  The subtracted
image is shown in Fig. \ref{fig:ngc4649}.

\subsection{NGC\,4696}

NGC\,4696 is the dominant galaxy of the Centaurus Cluster (Abell 3526)
and hosts the radio source PKS 1245-41.  The
radio emission has two clear lobes.
High frequency radio emission shows that the jets initially travel
east-west before both turning south \citep{Taylor02}.  The interaction of the
radio source with the surrounding
intra-cluster medium is strikingly clear and is discussed in detail by \citet{Sanders02,Fabian05,Taylor06a}.  

\subsection{NGC\,5044}

NGC\,5044 is the central member of a rich group \citep{Rickes04}). 
Our VLA A-configuration data detect a central point radio source.
The X-ray emission at the centre of this
group is disturbed \citep{Buote03a,Buote03b}.  \citet{Gastaldello09} cite evidence for a pair
of bubbles, one north and one south of the X-ray core.  Although there
is no high frequency radio emission associated with them, H$\alpha$ and [ N
  {\scshape ii} ] emission does appear to be correlated with 
depressions in the X-ray gas \citep{Gastaldello09,Caon00}.  More
recent GMRT\footnote{Giant Metrewave Radio Telescope} observation show
that the southern cavity is filled by radio plasma emitting at $610$ and
$235\mhz$ radio emission \citep{David09}.  

\subsection{NGC\,5846}

NGC\,5846 is one of a dominant galaxy pair (with NGC\,5850) of the
NGC\,5846 group.  
Our VLA A-configuration data detect a central point radio source.
The morphology of the X-ray emission is highly
suggestive of an AGN interaction, with a clear circular cavity to the
north-east of the core \citep{Trinchieri02}.  There is also a wedge to the south west that
is fainter than might otherwise be expected.  
Since no radio emission is associated with the X-ray cavities, here
we conservatively classify them as `possible' rather than `definite'.
Deeper, low frequency radio observations are required.

\section{Population Study}\label{sec:population}

\subsection{Prevalence of Radio Activity}\label{sec:pop:RAct}

Our sample contains the X-ray brightest, early-type (elliptical and
S0) galaxies in the local universe. The sample is drawn from a parent
catalogue of optically-selected, optical magnitude-limited galaxies that
is itself approximately 90 per cent complete. Using deep VLA
observations we have investigated the radio fluxes and morphologies of
our galaxies.  Only one of the 18 galaxies in the sample is not
detected at radio wavelengths (NGC\,499).  Thus, 94 per cent of our
sample have some level of radio activity at their cores.

The one galaxy for which no central radio source is detected,
NGC\,499, is an X-ray bright lenticular (E/S0) galaxy.  The upper
limits on the radio emission from this galaxy are stringent.  Compared
to all other galaxies, NGC\,499 is very deficient in its $1.5\ghz$ radio
luminosity.

Of the 17/18 galaxies with detected
radio emission, 10 (55 per cent) have extended radio lobes.  We note
that most of our observations are only at one frequency, and future
observations, especially at low frequencies, may detect extended
features in some sources which so far only exhibit a point source.  In
all cases (apart from IC\,310) 
where the radio morphology is clearly extended, there are features in
the X-ray emission which correlate with the radio lobes, although in
one case they are faint.  The converse is not quite true.  In total there
are 12 galaxies (67 per cent) which have disturbed X-ray morphologies, which in some
cases are clearly bubble-like.  These 12 include all nine
which have extended radio emission (excluding IC\,310).  The other
three (NGC\,533, NGC\,5044 and NGC\,5846) have bubble-like features in the X-ray emission, but
show no clear link to the high frequency radio morphology.

For the radio point sources, deeper and lower frequency
observations with the Long 
Wavelength Array (LWA, \citealp{Ellingson09}) or Low Frequency Array
(LOFAR, \citealp{deVos09}; both LOFAR and the LWA are currently under
construction.) would be useful in searching for past 
episodes of activity from the central AGN.

Unfortunately 4/18 of our galaxies are observed at
$4.9\ghz$\footnote{The VLA archive was searched to find the radio
  observations which showed the most detail for these galaxies, with
  the aim to find $1.4\ghz$ observations.  $4.9\ghz$ observations were
  only used when no suitable $1.4\ghz$ observations were found.}.  We
therefore note that the observations of these galaxies are therefore
slightly biased against detecting extended radio emission.  Hence, the
number of galaxies which have extended radio emission could be a lower
limit on the true number.

Only two of the eighteen galaxies in our sample, IC\,310 and
NGC\,4203, host strong X-ray emitting AGN.  However, the inclusion of
these two galaxies does not affect our results or conclusions significantly.

\subsection{The link between cool cores and radio activity}\label{sec:pop:CCores}

Recent observations of the action of AGN at the centres of groups and
clusters have alluded to a feedback process between the AGN and the
surrounding gas (see e.g. \citealp{Churazov02, Allen06, McNamara07}).
The cooling of the X-ray emitting gas provides the fuel for the AGN,
whose output is therefore regulated by the amount of gas cooling.
However, the output of the AGN prevents the gas from cooling as
rapidly as it otherwise would.  Estimates of the energy injected by
the AGN into the central regions of the clusters and galaxies appears
to match that required to prevent (excessive) cooling (e.g. \citealp{Birzan04,
  Dunn06,Rafferty06}).  However, the
details of the coupling of the injected energy to the intracluster
medium or galactic halo are not fully understood.
Attempts at modelling this self-regulated feedback have made some
progress over recent years (e.g. \citealp{Vernaleo06,
  Brueggen09}), but as yet there is no observational evidence of the
coupling mechanisms.  Hence, this picture of feedback is conceptually attractive,
although a number of details remain unclear.

\begin{figure}
\centering
\includegraphics[width=1.0 \columnwidth]{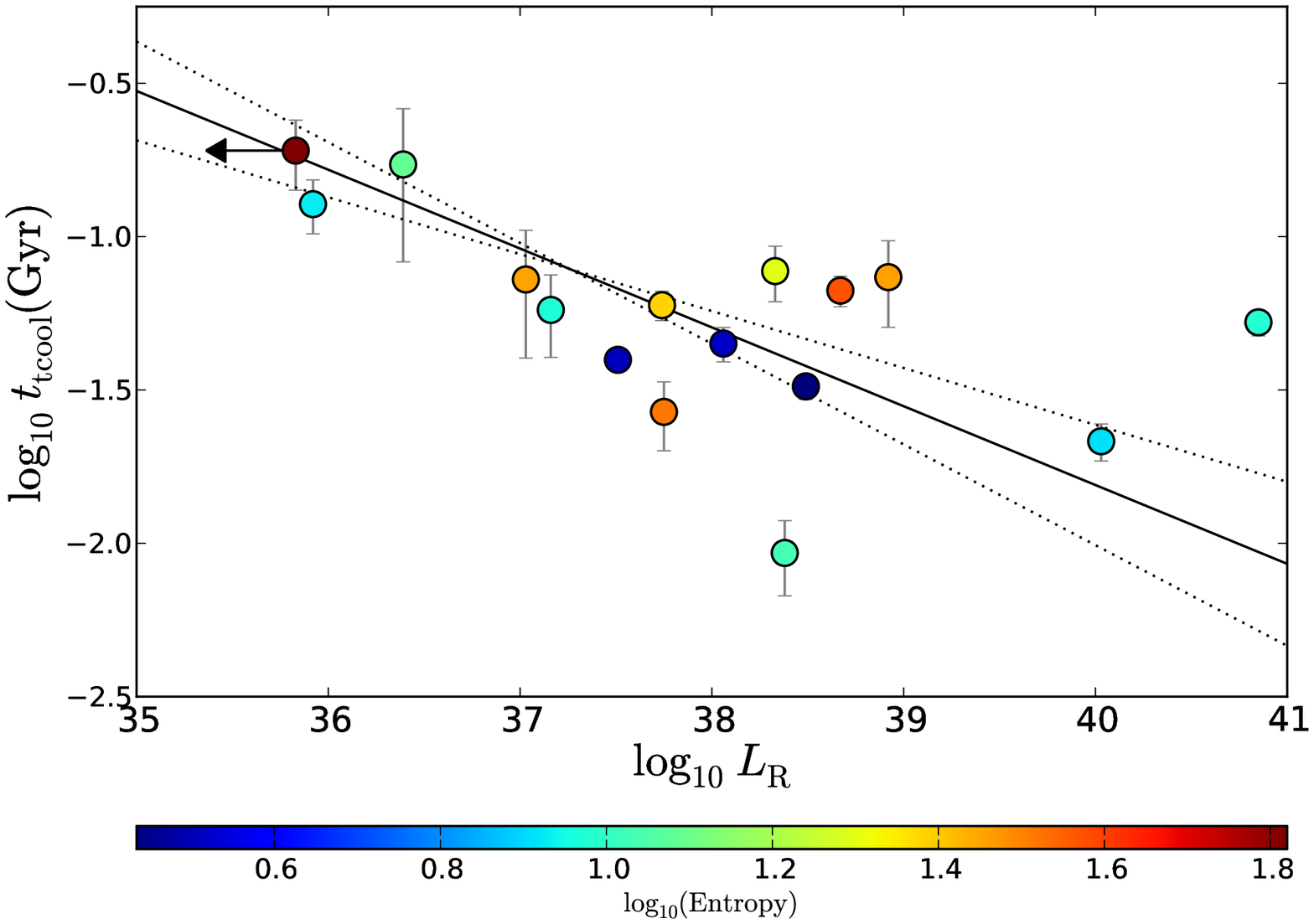}
\caption{\label{fig:tcool} The mass-weighted cooling time within
  $1\kpc$ against the $L_{\rm \nu_{\rm  R}}$ for all galaxies except
  IC\,310 and NGC\,4203.  The best fit lines
  shown are the BCES Bisector (solid) along with the errors on the Bisector (dotted) are
shown.  The colour scale shows the mass-weighted central
entropy within $1\kpc$ for the galaxies.}
\end{figure}

To test this scenario, we calculate the mean mass-weighted cooling
times within $1\kpc$ for all 
the galaxies and compare them to $L_{\rm R}$, the
monochromatic radio luminosity (see Fig. \ref{fig:tcool}).  Here,
$L_{\rm R}=\nu_{\rm  R}L_{\rm \nu_{\rm  R}}$ where $\nu_{\rm  R}$ and
$L_{\rm \nu_{\rm  R}}$ are the observing  
frequency and monochromatic radio luminosity as shown in Table
\ref{tab:radioL}.  The 
cooling times indicate how rapidly the gas is cooling onto the SMBH,
and hence give a rough measure of the fuelling rate of the AGN.  The radio
luminosity, as we include the large scale emission from the radio
lobes, gives a crude measure of the mechanical output of the AGN from the
relativistic jets.  For more precise calculations with a smaller
sample, see \citet{Allen06}.

In many of the X-ray observations of the galaxies in the sample a
central point source was identified and removed.  This creates a
minimum radius for which our annular deprojection can obtain values
for the cooling time.  For these systems, to estimate the cooling time
we fit simple power laws to the density and cooling time
profiles and extrapolate these inwards to obtain appropriate values
for the missing inner annular region.  Monte Carlo simulations are
used to estimate the final uncertainties in the results.

Setting aside the two galaxies for which the X-ray emission is dominated by nuclear sources
(IC\,310 and NGC\,4203) we observe (see Fig. \ref{fig:tcool}) a slight anti-correlation between the
radio luminosity and the mass-weighted cooling time, with a slope of
$-0.14$.  The BCES \citet{Akritas96} slope estimators are shown in
Table \ref{tab:bces}, and take into account uncertainties in both
parameters and the possible presence of intrinsic scatter.  Although  
the sample is small and the scatter is large, we do see indication of
a link between the central cooling 
times and the radio luminosity which is intriguing.

Our findings are consistent with the results of \citet{Allen06} and
lend further support to the idea that the X-ray gas fuels the central
AGN in these systems.

\begin{table}
\caption{BCES fits to $T_{\rm cool}$ vs $L_{\rm R}$\label{tab:bces}}
\begin{tabular}{lcccrccl}
\hline
\hline
Method & Slope & Intercept & Bootstrap Slope \\
\hline
     BCES(Y|X) & $-0.138 \pm  0.0515$  & $3.96 \pm  1.93$ &$-0.146 \pm  0.0631$ \\
     BCES(X|Y) & $-0.383 \pm  0.147 $  & $13.3 \pm  5.56$ &$-0.457 \pm  0.926$ \\
  BCES Bisector& $-0.257 \pm  0.0714$  & $8.47 \pm  2.66$ &$-0.275 \pm  0.123$ \\
BCES Orthogonal& $-0.142 \pm  0.0538$  & $4.14 \pm  2.07$ &$-0.151 \pm  0.0664$ \\
\hline
\end{tabular}
\begin{quote}
The slopes from the BCES estimator are also shown in Fig. \ref{fig:tcool}.
\end{quote}
\end{table}

Even in this comparatively small sample, the link between extended
radio emission and features in the X-ray gas is not totally clear.
Although all sources which have extended radio emission have features
in the X-ray gas, the converse is not true.  In studying the effect of
AGN on their surroundings, using {\it only} the X-ray morphology could
give misleading results.  The selection involved in finding
AGN-inflated cavities are not presently well understood
\citep{Nulsen06,Birzan09}.  The fraction of galaxies 
showing AGN inflated cavities 
in this sample is around twice what \citet{Nulsen06} report.  The high level
of core radio detections in the sample presented here indicate that,
{\it (a)} the fraction of 
AGN currently injecting mechanical energy into their surrounding could 
be much higher than previously thought, and {\it (b)} 
the criteria for an AGN to be ``active'' need to be quantified.  

The AGN in the elliptical galaxies with detected cavities in
\citet{Nulsen06} appear to be injecting more energy than required to
offset the cooling of the X-ray gas.  This may result if the cavities
for an average elliptical are very difficult to detect, and so only
exceptional (overpowered) outbursts are (usually) observed.  For lower
mass systems (groups and ellipticals), the overall level of
(mechanical) energy injection into their surrounding environments
remains incompletely understood.

\subsection{Radio-to-X-ray flux ratios}\label{sec:pop:RL}

We have determined integrated radio-to-X-ray flux ratios for the
galaxies. The measured quantity, $\mathcal{GR}_{\rm X}$, follows the
radio-to-X-ray loudness parameter for nuclear AGN emission,
$\mathcal{R}_{\rm X}$, determined by \citet{Terashima03}.  We write
\begin{equation}
\mathcal{GR}_{\rm X}=\nu_{\rm  R} L_{\rm \nu_{\rm  R}}/L_{\rm X}.
\end{equation}
\noindent where $L_{\rm X}$
is the $0.1-2.4\kev$ band X-ray luminosity. \citet{Terashima03} note 
that the boundary between radio-loud and radio-quiet 
nuclear sources occurs at $\log_{10} (\mathcal{R}_{\rm X})=-4.5$. 

\begin{figure}
\centering
\includegraphics[width=1.0 \columnwidth]{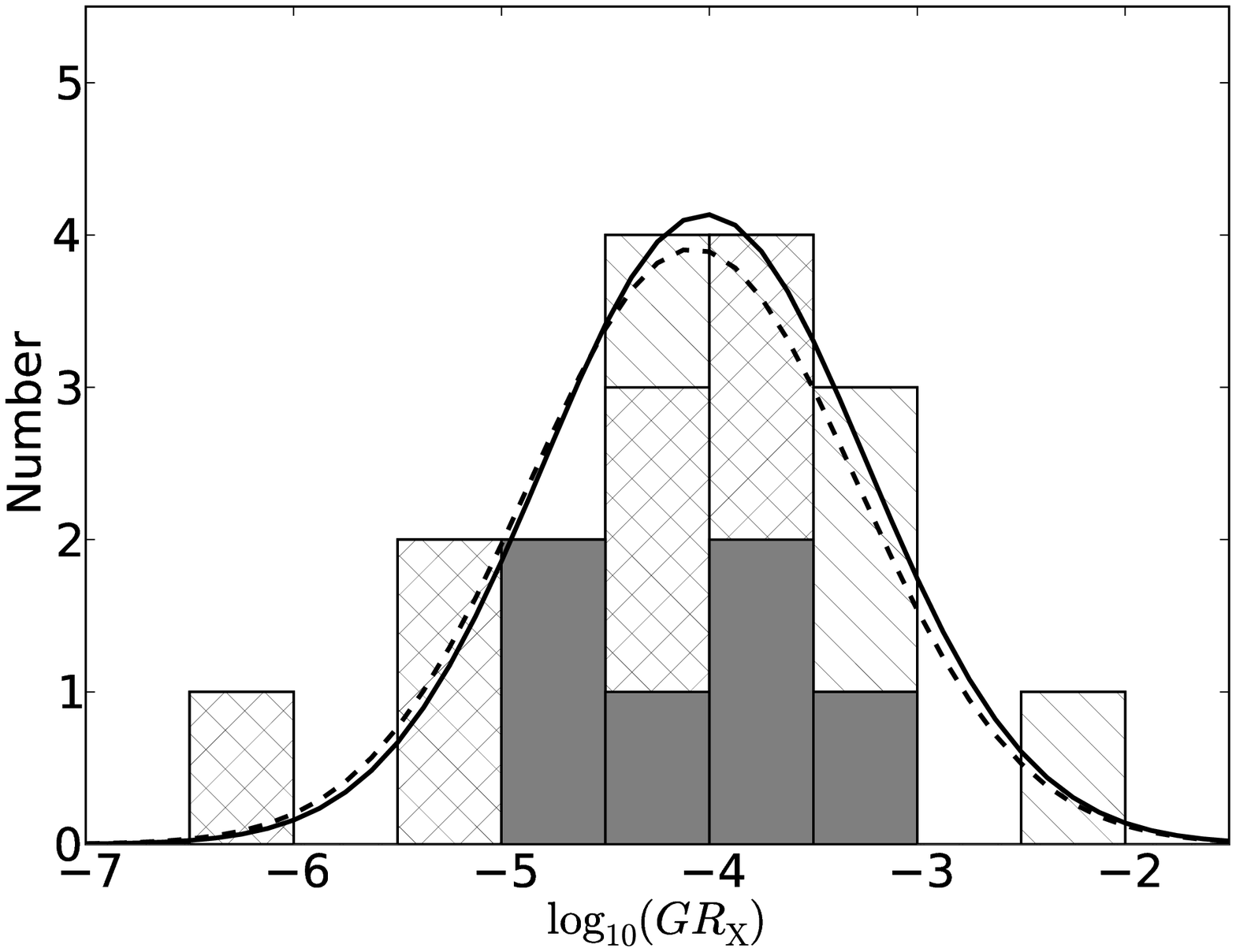}
\caption{\label{fig:RL} The distribution of $\log(\mathcal{GR}_X)$ for the
sample of 17 galaxies with detected radio emission.  The hashed bars
are the group dominant- , the cross-hashed are the cluster dominant-
and the solid bars are the non-dominant galaxies.  The two fitted
curves are for a least-squares (dashed) and a downhill simplex (solid)
Gaussian model fit to the total distribution.  IC\,310 and NGC\,4203 are included in this plot. }
\end{figure}

The distribution of
$\log_{10}(\mathcal{GR}_{\rm X})$ is shown in Fig. \ref{fig:RL}, 
with the data detailed
in Table \ref{tab:radioL}.  As can be seen from Fig. \ref{fig:RL}, 
there is a large spread in
$\log(\mathcal{GR}_X)$. Also shown are the results of fitting a Gaussian
model to the distribution: the peak of the Gaussian lies 
at $\log_{10}(\mathcal{GR}_{\rm
  X})=-4.02\ (-4.08)$ with a variance of $0.78\ (0.79)$ for a
downhill simplex (least squares) fit.  

We have also calculated 
$\mathcal{GR}_{\rm X,\,NVSS}$ values 
based on the radio flux densities determined from the 
NVSS Survey \citep{Condon88}.  We show the monochromatic radio
luminosities from the NVSS, $L_{\rm R,\,NVSS}$, along with the
$\mathcal{GR}_{\rm X,\,NVSS}$ in Table \ref{tab:radioL}. The NVSS flux
densities were all  
measured at $1.4\ghz$ using a more compact
configuration of the VLA. More faint emission is included at the price
of the lower spatial resolution.  There are some differences between
the values of $\mathcal{GR}_{\rm X}$ calculated from the high spatial
resolution VLA observations and those from the NVSS. However, 
all but one (IC\,310) differ by less than a factor of two. 
All galaxies in our sample have NVSS detections.  See
Section \ref{sec:extended} for more on the NVSS properties.

\begin{table*}
\caption{Radio Loudness\label{tab:radioL}}
\begin{tabular}{lcccccccc}
\hline
\hline
Source&Type&Frequency&$L_{\rm X}$&$S_{\rm \nu R}$&$L_{\rm R}$&${\mathcal{GR}}_{X}$ &$L_{\rm R,\,NVSS}$&${\mathcal{GR}}_{\rm X,\,NVSS}$ \\
&&(\ghz)&$(\ergps)$&$(\ergpspHz)$&$(\ergps)$&&$(\ergps)$&\\
\hline
IC\,1860  & P & 1.365 &  $5.89\times 10^{42}$ &  $1.39\times 10^{-25}$ &$2.12\times 10^{39}$ & $3.60\times 10^{-5}$ &$2.86\times 10^{38}$ & $4.86\times 10^{-5}$ \\
NGC\,499  & P & 1.365 &  $2.24\times 10^{42}$ &  - & - & - & - & - \\	        		 	            					    
NGC\,507  & E & 1.525 &  $5.75\times 10^{42}$ &  $8.74\times 10^{-25}$ &$8.27\times 10^{38}$ & $1.44\times 10^{-4}$ &$5.36\times 10^{38}$ & $9.31\times 10^{-5}$ \\	
NGC\,533  & P & 4.885 &  $1.95\times 10^{42}$ &  $8.91\times 10^{-26}$ &$2.42\times 10^{38}$ & $1.24\times 10^{-4}$ &$2.23\times 10^{38}$ & $1.14\times 10^{-4}$ \\
NGC\,708  & E & 1.365 &  $1.23\times 10^{43}$ &  $8.25\times 10^{-25}$ &$4.72\times 10^{38}$ & $3.83\times 10^{-5}$ &$3.85\times 10^{38}$ & $3.13\times 10^{-5}$ \\
NGC\,1399 & E & 1.465 &  $4.90\times 10^{41}$ &  $4.63\times 10^{-24}$ &$3.06\times 10^{38}$ & $6.25\times 10^{-4}$ &$1.31\times 10^{38}$ & $2.68\times 10^{-4}$ \\
NGC\,1404 & P & 1.365 &  $1.78\times 10^{41}$ &  - & - & -  							    &$2.46\times 10^{36}$ & $1.38\times 10^{-5}$ \\	    
NGC\,1550 & E & 1.365 &  $7.24\times 10^{42}$ &  $1.27\times 10^{-25}$ &$5.07\times 10^{37}$ & $7.73\times 10^{-6}$ &$7.51\times 10^{37}$ & $1.04\times 10^{-5}$ \\
NGC\,4203 & P & 4.885 &  $1.74\times 10^{41}$ &  $1.80\times 10^{-25}$ &$3.18\times 10^{37}$ & $1.83\times 10^{-4}$ &$3.09\times 10^{36}$ & $1.78\times 10^{-5}$ \\
NGC\,4406 & P & 4.885 &  $1.29\times 10^{42}$ &  $4.91\times 10^{-27}$ &$8.35\times 10^{35}$ & $6.48\times 10^{-7}$ & - & - \\
NGC\,4472 & E & 1.489 &  $3.09\times 10^{41}$ &  $2.23\times 10^{-24}$ &$1.16\times 10^{38}$ & $3.74\times 10^{-4}$ &$1.07\times 10^{38}$ & $3.47\times 10^{-4}$ \\	
NGC\,4486 & E & 1.435 &  $1.02\times 10^{43}$ &  $8.73\times 10^{-22}$ &$7.01\times 10^{40}$ & $6.85\times 10^{-3}$ &$6.75\times 10^{40}$ & $6.60\times 10^{-3}$ \\  
NGC\,4636 & E & 1.425 &  $4.47\times 10^{41}$ &  $6.45\times 10^{-25}$ &$3.20\times 10^{37}$ & $7.16\times 10^{-5}$ &$3.79\times 10^{37}$ & $8.49\times 10^{-5}$ \\	
NGC\,4649 & E & 1.465 &  $2.19\times 10^{41}$ &  $2.82\times 10^{-25}$ &$1.44\times 10^{37}$ & $6.58\times 10^{-5}$ &$1.42\times 10^{37}$ & $6.48\times 10^{-5}$ \\
NGC\,4696 & E & 1.565 &  $1.95\times 10^{43}$ &  $3.64\times 10^{-23}$ &$1.07\times 10^{40}$ & $5.50\times 10^{-4}$ & - & - \\  
NGC\,5044 & P & 1.465 &  $6.31\times 10^{42}$ &  $2.99\times 10^{-25}$ &$5.49\times 10^{37}$ & $8.70\times 10^{-6}$ &$6.09\times 10^{37}$ & $9.65\times 10^{-6}$ \\
NGC\,5846 & P & 1.465 &  $5.13\times 10^{41}$ &  $1.02\times 10^{-25}$ &$1.08\times 10^{37}$ & $2.10\times 10^{-5}$ &$2.12\times 10^{37}$ & $4.13\times 10^{-5}$ \\
\hline
IC\,310   & E & 1.465 &  $3.98\times 10^{42}$ &  $1.50\times 10^{-24}$ &$1.21\times 10^{39}$ & $3.05\times 10^{-4}$ &$1.30\times 10^{39}$ & $3.26\times 10^{-4}$\\
NGC\,4203 & P & 1.465 &  $1.74\times 10^{41}$ &  $7.86\times 10^{-26}$ &$4.15\times 10^{36}$ & $3.10\times 10^{-4}$ &$3.09\times 10^{36}$ & $1.78\times 10^{-5}$ \\
\hline
\end{tabular}
\begin{quote}
The Types show whether the radio morphology is P-point like or
E-extended.  The $L_{\rm R}$ are calculated at the frequency shown in the table,
without any scaling to a common frequency.  We also show the $L_{\rm
  R,\,NVSS}$ calculated from the NVSS, which are all at the same
frequency.  We separate IC\,310 and NGC\,4203 which appear to be a different
class of source as the other galaxies.
\end{quote}
\end{table*}

Our sample contains four dominant cluster galaxies (NGC\,708,
NGC\,1399, NGC\,4486, NGC\,4696), seven dominant group galaxies
(IC\,1860, NGC\,507, NGC\,533, NGC\,1550, NGC\,4406, NGC\,4649,
NGC\,5044) and seven galaxies (IC\,310, NGC\,499, NGC\,1404, 
NGC\,4203, NGC\,4472, NGC\,4636, NGC\,5846) that are $not$ the
dominant galaxy of a group or cluster.  We see no significant 
difference in the $\mathcal{GR}_{\rm X}$ values for dominant 
and non-dominant galaxies.

\subsection{Prevalence of Radio-Loud AGN}

There is a perception that the fraction of AGN in the general
population of galaxies is small, even though SMBH are present in most
moderately sized galaxies.  Part of this perception is
historical: early studies only found AGN
fractions of a few per cent \citep{Dressler85,Huchra92}.  However, as
the data have improved, this fraction has 
increased \citep{Carter01,Miller03,Santra07} with nuclear
activity being found in galaxies that would not, classically, have
been classed as AGN.  For example, in the Palomar sample of nearby
bright galaxies (magnitude limited), \citet{Nagar05} and \citet{Filho06} used high
resolution radio imaging to show that
at least a quarter of galaxies, and probably more, are AGN.  However, many of these AGN are
low luminosity.  In another study, a combined optical and X-ray study of Elliptical
galaxies in the Virgo Cluster find X-ray signatures
of AGN in the cores of 49-87 per cent target galaxies with stellar
masses $>10^{10}M_{\odot}$ \citep{Gallo08}.  At lower galaxy masses the fraction falls to
3-44 per cent.  This decline in the AGN fraction with decreasing
galaxy mass has been seen at the cluster and group level by \citet{Best05}.   

\begin{figure}
\centering
\includegraphics[width=1.0 \columnwidth]{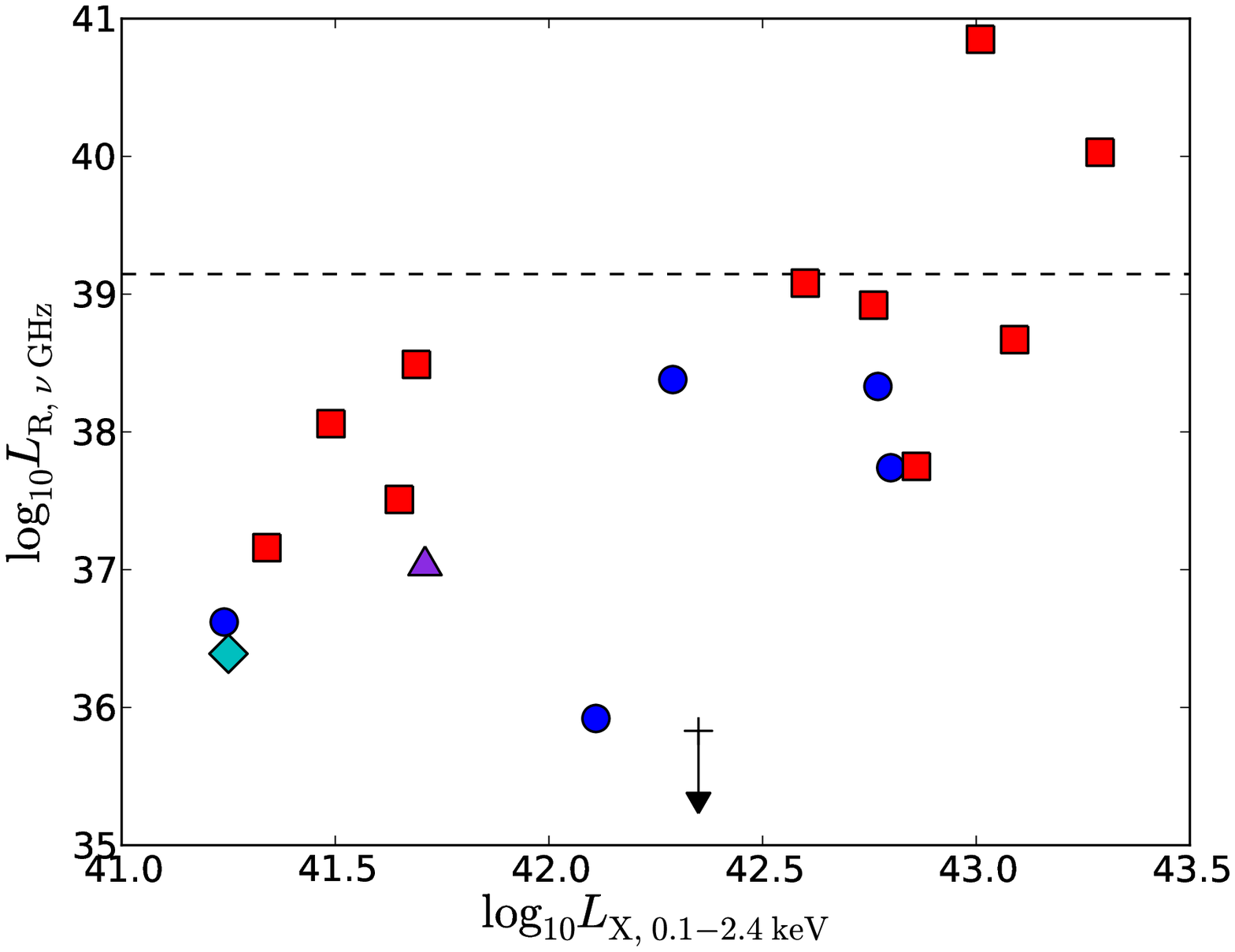}
\caption{\label{fig:Radio} The radio luminosity ($L_{\rm R, \nu}$) as a
  function of the X-ray luminosity ($L_{\rm X, 0.1-2.4 keV}$) along
  with a dashed line showing radio loudness cut-off used by 
  \citet{Best05}.  The red squares are extended radio
sources and the blue circles are radio point sources.  The purple triangle is
NGC\,5846 which although being a point source has clear indications of
interactions in the X-ray emission.  The cyan diamond is NGC\,1404
using NVSS data.  The upper limit for NGC\,499 is also shown.  The
limitations of using a cut-off in the $L_{1.4\ghz}$ are clear.}
\end{figure}

In determining the fraction of radio loud galaxies present in a sample
of SDSS galaxies \citet{Best05} use a limit for radio loudness of
$L_{1.4\ghz}>10^{23}\rm{W~Hz^{-1}}$.  We converted this limit to a
luminosity limit at the frequency of the NVSS catalogue ($1.5\ghz$).  Therefore, any galaxy
with $L_{\rm R,\,NVSS}>1.4\times 10^{39} \ergps$ would be classed as
radio-loud in \citet{Best05}.  In our sample only NGC\,4486 is above this
limit; however NGC\,4696 would also be if it was in the NVSS
catalogue, and IC\,310 falls directly on the boundary.  This gives a
radio loud fraction of 11 per cent, (2/18) and 17 per cent (3/18) if including
IC\,310.  We show the distribution of radio luminosity with X-ray
luminosity in Fig. \ref{fig:Radio}, along with the threshold value
for radio-loudness from \citet{Best05}.   

This radio loud fraction is
much lower than the {\it radio detected} fraction in our sample (17/18).  Although using a cut-off in
the $L_{1.4\ghz}$ ensures selecting AGN radio activity out to large
distances, the results from this study shows that it does not include
a large number of AGN with low radio luminosities.  Of the AGN which
are not included when using a radio luminosity cut-off, a significant
fraction have extended radio emission, though others are just radio
cores.  To obtain a clearer picture of the level of radio activity
within a sample of galaxies, the
radio-to-X-ray flux ratio (Section \ref{sec:pop:RL}) of a galaxy
can be used.   The
two galaxies which are classed as radio 
loud from the radio luminosity limit both have extended radio lobes.
However a further 8 (including 
IC\,310) also have extended radio emission, indicating active mechanical
energy injection into their surroundings.  Therefore the radio
luminosity limit results in an incomplete picture of the level of
energy injection into the surroundings of elliptical galaxies.  

Although our sample selects massive galaxies which are X-ray bright,
we are not purposefully selecting AGN host galaxies; central point
sources typically account for only a very small fraction of the total
X-ray flux.  Our results indicate that AGN activity is common in most massive galaxies;
massive galaxies being the ones which are more likely to have and be
able to retain a halo of X-ray gas.

In a large sample of clusters and groups of galaxies, \citet{Best07} find that
brightest group and cluster galaxies are more likely to host a radio
loud AGN than a field galaxy of an equivalent mass.  This increased
likelihood also extends to galaxies which are close to the cluster
centre.  Not all of our targets are dominant galaxies.  Most are in
clusters or groups.  The two most radio loud sources in our study are
the brightest galaxies of the Virgo and Centaurus Clusters.

\subsection{Particle Content}\label{sec:pop:kf}

Here we follow the analysis in \citet{Dunn04, Dunn05} and more
recently \citet{Croston05,DeYoung06,Shurkin07} and \citet{Croston08}
in studying the particle content of the radio lobes.  Under the
assumption that the relativistic radio-emitting plasma is in pressure
balance with the X-ray gas, we are able to determine the energy, and
hence particle content of the radio plasma.  

Measurements of the synchrotron emission from the radio lobes can be
used to calculate the energy contained within the relativistic
electrons present in the lobes.  \citet{Fabian02} studied the lobes of
3C~84 in the Perseus cluster.  Subsequently \citet{Dunn04, Dunn05}
investigated a larger number of radio sources in galaxy clusters.  For
a more detailed description of the method see e.g. \citet{Dunn04}.

For a continuous synchrotron spectrum with a single spectral
index, $\alpha$, between $\nu_1=10\mhz$ and $\nu_2=10\ghz$, the energy
in relativistic elections is
\begin{eqnarray}\label{eqn:Ee}
E_{\rm e}&\propto&
\frac{S_{\nu}}{\nu}\frac{\nu_2^{0.5+\alpha}-\nu_1^{0.5+\alpha}}{\alpha+0.5}
B^{-3/2} \erg\\
&\approx& aB^{-3/2}
\end{eqnarray}
\noindent Also taking into account the energy within the magnetic field, the total energy in
the lobes is 
\begin{equation}
E_{\rm tot}=kE_{\rm e} + Vf\frac{B^2}{8\pi},
\end{equation}
\noindent where $k$ accounts for any other particles present in the lobe which
are not accounted for by the simplistic model spectrum.
$V=\frac{4}{3}\pi R_{\rm l}R_{\rm w}^2$ is the
volume of the bubble, where $R_{\rm l}$ and $R_{\rm w}$ are the bubble
radii along and perpendicular to the jet axis respectively.  $f$ is
the volume filling fraction of the relativistic plasma.

The magnetic field present inside the bubble is estimated by comparing
the synchrotron cooling time of the plasma to the age of the lobe.
The latter can be estimated from the sound-speed expansion
timescale\footnote{Since the radio lobes of interest are still
  attached to the active radio jets, this is the most sensible, easy
  to calculate, timescale.}.  No strong shocks are seen in the X-ray
gas, so sound speed expansion timescale gives a
lower limit on the age of the bubbles, and hence an upper limit on the
magnetic field.

We obtain upper limits on the ratio $k/f$, 
\begin{equation}
\frac{k}{f} \leq \left( P_{\rm th}-\frac{B^2}{8\pi}\right)\frac{3V}{a}B^{3/2}
\end{equation}
\noindent for all galaxies with extended radio emission, where the
$3V$ arises from the energy density of the relativistic particles.  If $k/f=1$,
i.e. $\log_{10}(k/f)=0$, then the lobe is filled with a purely
electron-positron plasma with emission only in the range $10\mhz$ to
$10\ghz$.  Under the assumption $f\sim 1$, then if $k/f>1$, $k>1$,
which implies that there are ``extra'' particles required for the lobe
to be in pressure balance with its surroundings.  These could be
thermal protons, mixed into the relativistic plasma as the jet travels
out from the AGN, or they could be electrons which radiate out of the
assumed region. Since we assume a simple power-law 
slope for the spectral index,
any large deviation at low (and hence unobserved) radio frequencies
will also change the calculated $k/f$.  We have also determined the value of
$k/f$ if the sources were to be in equipartition.  However in none of
the radio bubbles does equipartition between the relativistic
particles and the magnetic field lead to pressure balance between the
radio and X-ray plasmas.

\begin{table*}
\caption{$k/f$ Results\label{tab:kf}}
\begin{tabular}{lccccccccccc}
\hline
\hline
Galaxy & Lobe & $R_{\rm l}$&$R_{\rm w}$&$R_{\rm Dist}$&$S$ &$\alpha$&$\log(B)$&
$\log(k/f_{\rm eqptn})$ & $\log(k/f_{\rm sound})$ & $\log(\nu_{\rm
  min})$ & $\log(k/f_{\nu_{\rm min}})$ \\
&&(kpc)&(kpc)&(kpc)&(Jy)\\
\hline
NGC507 & E & 10.70 & 15.20 & 14.90 &  0.027 &$-1.40\pm 0.20$& $-5.09\pm 0.02$&$ 2.83\pm 0.34$&$ 2.77\pm 0.34$&$ 4.04\pm 0.95$&$-0.00\pm 0.02$\\ 
NGC507 & W & 11.00 & 14.30 & 12.20 &  0.048 &$-1.40\pm 0.20$& $-5.10\pm 0.02$&$ 2.54\pm 0.35$&$ 2.47\pm 0.34$&$ 4.37\pm 0.87$&$-0.00\pm 0.02$\\ 
NGC708 & E &  4.93 &  2.72 &  5.78 &  0.014 &$-0.60\pm 0.20$& $-4.86\pm 0.03$&$ 3.52\pm 0.17$&$ 3.34\pm 0.16$&$ 1.58\pm 0.03$&$ 2.68\pm 0.78$\\ 
NGC708 & W &  4.75 &  2.82 &  4.75 &  0.017 &$-0.60\pm 0.20$& $-4.85\pm 0.03$&$ 3.44\pm 0.18$&$ 3.28\pm 0.18$&$ 1.59\pm 0.03$&$ 2.61\pm 0.78$\\ 
NGC1399 & N &  2.88 &  1.94 &  8.39 &  0.11 &$-0.90\pm 0.10$& $-4.73\pm 0.03$&$ 2.56\pm 0.17$&$ 2.62\pm 0.17$&$ 1.67\pm 0.03$&$ 0.01\pm 0.02$\\ 
NGC1399 & S &  3.96 &  2.50 &  8.42 &  0.13 &$-0.90\pm 0.10$& $-4.81\pm 0.03$&$ 2.60\pm 0.18$&$ 2.65\pm 0.17$&$ 1.63\pm 0.02$&$ 0.01\pm 0.01$\\ 
NGC1550 & E &  4.34 &  6.37 &  5.60 &  0.0035 &$-1.00\pm 0.50$& $-4.83\pm 0.03$&$ 6.83\pm 0.58$&$ 6.63\pm 0.57$&$ 1.62\pm 0.03$&$-0.01\pm 0.02$\\ 
NGC1550 & W &  7.34 &  4.55 & 10.70 &  0.0029 &$-1.00\pm 0.50$& $-4.97\pm 0.03$&$ 6.51\pm 0.57$&$ 6.25\pm 0.57$&$ 1.46\pm 0.03$&$-0.01\pm 0.02$\\ 
NGC4472 & E &  2.24 &  2.24 &  4.81 &  0.029 &$-0.80\pm 0.10$& $-4.68\pm 0.03$&$ 3.65\pm 0.15$&$ 3.70\pm 0.15$&$ 1.72\pm 0.03$&$ 2.08\pm 0.60$\\ 
NGC4472 & W &  3.07 &  1.83 &  2.82 &  0.038 &$-0.80\pm 0.10$& $-4.77\pm 0.03$&$ 3.27\pm 0.14$&$ 3.30\pm 0.14$&$ 1.66\pm 0.02$&$ 1.66\pm 0.61$\\ 
NGC4486 & J &  1.46 &  0.91 &  1.49 & 35.8 &$-0.60\pm 0.20$& $-4.47\pm 0.02$&$ 1.13\pm 0.22$&$ 0.97\pm 0.22$&$ 1.96\pm 0.02$&$ 0.01\pm 0.02$\\ 
NGC4486 & CJ &  1.29 &  1.29 &  1.48 & 56.5 &$-0.60\pm 0.20$& $-4.44\pm 0.02$&$ 1.23\pm 0.16$&$ 1.11\pm 0.16$&$ 1.99\pm 0.03$&$ 0.01\pm 0.01$\\ 
NGC4636 & NE &  0.71 &  0.41 &  0.82 &  0.028 &$-0.60\pm 0.20$& $-4.40\pm 0.02$&$ 2.43\pm 0.18$&$ 2.59\pm 0.18$&$ 2.04\pm 0.02$&$ 0.01\pm 0.02$\\ 
NGC4636 & SW &  0.77 &  0.39 &  0.86 &  0.0185 &$-0.60\pm 0.20$& $-4.42\pm 0.01$&$ 2.61\pm 0.16$&$ 2.76\pm 0.16$&$ 2.02\pm 0.01$&$ 0.01\pm 0.02$\\ 
NGC4649 & N &  0.51 &  0.49 &  0.84 &  0.0029 &$-0.60\pm 0.20$& $-4.25\pm 0.02$&$ 4.13\pm 0.16$&$ 4.26\pm 0.16$&$ 2.17\pm 0.01$&$ 3.64\pm 0.70$\\ 
NGC4649 & S &  0.65 &  0.42 &  1.03 &  0.0019 &$-0.70\pm 0.20$& $-4.32\pm 0.02$&$ 3.93\pm 0.22$&$ 4.06\pm 0.21$&$ 2.12\pm 0.02$&$ 3.04\pm 0.95$\\ 
NGC4696 & E &  2.83 &  1.75 &  2.32 &  1.3 &$-0.75\pm 0.20$& $-4.72\pm 0.02$&$ 1.53\pm 0.25$&$ 1.40\pm 0.25$&$ 1.68\pm 0.03$&$ 0.01\pm 0.01$\\ 
NGC4696 & W &  2.88 &  1.96 &  3.24 &  0.65 &$-0.75\pm 0.20$& $-4.73\pm 0.02$&$ 1.94\pm 0.24$&$ 1.81\pm 0.24$&$ 1.68\pm 0.03$&$ 0.01\pm 0.01$\\ 
\hline
\end{tabular}
\begin{quote}
$R_{\rm l}$ and $R_{\rm w}$ are the dimensions of the bubbles with
  $R_{\rm dist}$ their separation from the AGN.  $k/f_{\rm eq^n}$ is
  the equipartition value of $k/f$ and $k/f_{\rm sound}$ is the value
  as estimated from the sound speed expansion timescale.  $\nu_{\rm
    min}$ is the lower cut off to the radio spectrum required for
  $k/f=1$ if it can be reached before the cyclotron frequency.
  $k/f_{\rm \nu_{min}}$ is the value of $k/f$ at that frequency.  The
  bubble descriptors are N-North, S-South, E-East, W-West, J-Jet,
  CJ-Counter Jet.
The spectral indices have been taken from the following; NGC\,507
\citep{Colla75}, NGC\,708 \citep{Hardcastle03}, NGC\,1399
\citep{Ekers89}, NGC\,4649 \citep{Stanger86}, NGC\,4486
\citep{Hines89}, NGC\,4636 \citep{Stanger86}, NGC\,4472
\citep{Vollmer04}, NGC\,4696 \citep{Fabian05}.

\end{quote}
\end{table*}

\begin{figure}
\centering
\includegraphics[width=1.0 \columnwidth]{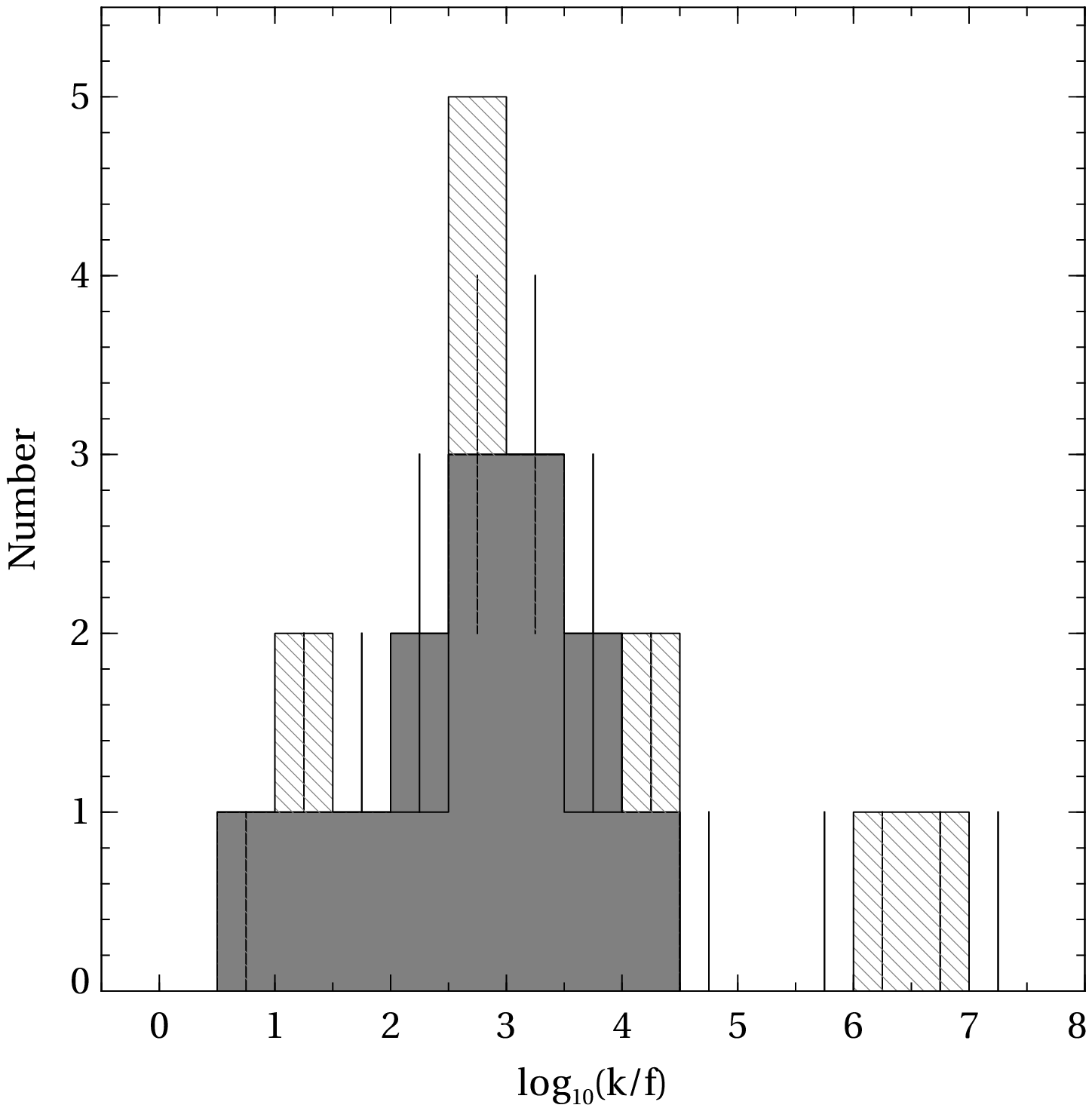}
\caption{\label{fig:KF} The distribution of the upper limits on $k/f$ for the
16 galaxies which have extended radio emission.  We show both a
standard binning (hashed) and a Monte-Carlo routine which was used to
account for the large uncertainties of the $k/f$ values (solid).  The
errorbars are from the Monte-Carlo routine and are for the solid bars.}
\end{figure}

As can be seen from Table \ref{tab:kf}, our sample 
exhibits a large range of $k/f$ upper limits.  The distribution of 
values is also shown in Fig. \ref{fig:KF}. (We use a Monte Carlo
algorithm to estimate the uncertainties in the upper limits). 
The distribution of $k/f$ is broad, in agreement with the 
conclusions of  e.g. \citet{Dunn05, Birzan07}.  

\subsubsection{Low Frequency Radio Spectrum}

The value of $k/f$ is expected to increase as a plasma ages with
synchrotron losses, causing the average energy of the particles to
drop. Electrons with low energies will emit at frequencies below the observed energy
range and act as ``extra particles'' in the model.  Since bubbles will
only appear `young' for a short time, most will be seen when $k/f$ is
large.  Eventually the radio emission will drop below detection
limits, leading to an upper limit to the values of $k/f$ estimated
from GHz radio data.  There is a suggestion from Fig.  \ref{fig:KF},
that the distribution peaks at around $10^3$, but with asymmetric tails
either side.

We have investigated whether, by modifying the spectrum to include
additional low frequency radio emission from low energy particles
(extending $\nu_1=\nu_{\rm min}<10\mhz$), it is possible to achieve
$k/f$ equal to one.  The minimum possible radio frequency is the
cyclotron frequency as estimated from the magnetic field estimates.
For around half of the bubbles $k/f=1$ is achieved before the
frequency limit is reached.  The bubbles for which $k/f=1$ is not
achievable are those which have the largest $k/f$ values to start
with.

In some of the galaxies in our sample, there is no extended $1.4\ghz$
radio emission, even when there are features in the X-ray emission
which point towards an interaction (e.g. NGC\,5044 and NGC\,5846).
Studies which have included low frequency radio emission
(e.g. \citealp{Dunn05, Birzan08}) obtain $k/f$ values which are
larger than the values from $\ghz$ radio emission.  However, the
increase in $k/f$ for ghost bubbles, regardless of the type of radio
emission is currently detected from them, is at least in part an effect of not being
able to sample the full radio spectrum of the lobes.  

As the electrons age, the synchrotron radio flux in the $\ghz$ band
falls.  Therefore more ``extra particles'' are required for pressure
balance, resulting in a higher value calculated for $k/f$ (assuming $f=1$).
However, the aged synchrotron electrons are still present in the lobe
and still supporting it, but radiating at much lower frequencies.
Therefore a measure of $k/f$ at $\ghz$ frequencies does not fully
describe the $k/f$ of the complete lobe.  Observations with the new
low frequency radio telescopes (LOFAR, LWA or the Murchison Widefield
Array \citealp{Lonsdale07}) will uncover the spectral shape at low
frequencies and so allow further investigations into the detected and
the presumed cavities.

\subsubsection{Possible Biases}

In the calculations outlined above, we have used the sound speed
timescale to estimate the age of the bubbles, and compared this age
to the synchrotron cooling time of the plasma to estimate the magnetic
field.  These are not the only possibilities available when
calculating the $k/f$ for the cavities.  

Older cavities have been described by a buoyancy timescale, the time
taken to rise buoyantly in the potential well of the galaxy \citep{Churazov00}.  This
buoyant rise takes place after the initial expansion, and separates
the cavity from the galactic nucleus.  In all of the cavities studied
here, they are still attached to the nucleus, with on the whole, radio
emission also connected to the radio core.  The buoyancy timescale is therefore unlikely
to be a reasonable description for the age of the cavities.

Another timescale of interest is the time for the intra-galactic medium to refill the
displaced volume \citep{McNamara00}.  However, this is also most applicable to
older cavities which have detached from the nucleus and risen up, with
the IGM flowing around behind the cavity.

No {\it strong} shocks are seen around the cavities at the
centres of nearby clusters of galaxies (e.g. Perseus, A2052), the
expansion of these bubbles is not currently highly
supersonic.  Although fed by a relativistic jet, and hence at some
point expand supersonically, on average the sound speed is an
appropriate value to calculate the age of the cavities from.  We
assume that this is also the case in these, in some cases lower-power,
cavity systems.

To obtain an estimate on the magnetic field of the relativistic
plasma, we compare the synchrotron cooling time to the age of the
cavity.  Using the observation that GHz radio emission is detected, we
conclude that the synchrotron cooling time must be longer than the age
of the cavity.  However, we are unable to say how much longer; we can
only place a lower bound on the synchrotron cooling time, and hence an
upper bound on the magnetic field strength.  Future low-frequency
radio observations will help in determining the spectral shape of the
radio emission, and hence improve the estimates on the magnetic field
strength, and hence the age of the bubbles.

Improving the age estimates of the bubbles may be possible with
increasingly accurate simulations of the behaviour for the evolution
of radio lobes in clusters of galaxies.  For older bubbles, a
combination of the sound speed expansion and the buoyancy rise time
may be the best current description for the ages.

\section{Extended Sample}\label{sec:extended}

We have attempted to extend our study to lower X-ray and radio
fluxes. Dropping the flux cut to $1\times 10^{-12}\ergpspcmsq$
increases the sample from 18 galaxies to 42.   Of these 42 galaxies, 25 (60 per
cent) have their $L_X$ values determined from follow-up pointed observations
\citep{OSullivan01}.  In order to
limit our study to the most massive elliptical and
S0 galaxies, we introduce a luminosity cut along with the flux cut.
Limiting the galaxies to those with $L_X>10^{41.0}\ergps$ ensures we
do not select very nearby low mass galaxies.  

To determine the radio properties of these galaxies we use the NVSS
survey at $1.4\ghz$ for the northern hemisphere, and the Sydney University Molonglo
Sky Survey (SUMSS, \citealp{Bock99,Mauch03}) at $843~{\rm MHz}$ for the southern
hemisphere.  These surveys, albeit at slightly different frequencies,
are comparable in terms of their sensitivity and resolution.
Using these, we are able to determine whether the galaxy hosts a radio
source.  We perform a search using a radius of 30 arcsec in NVSS and
then a 60 arcsec search for those galaxies where nothing is found.  We
use the SUMSS postage stamp cutouts to check those galaxies which are out of
the NVSS survey area.  

\begin{figure}
\centering
\includegraphics[width=0.49\textwidth]{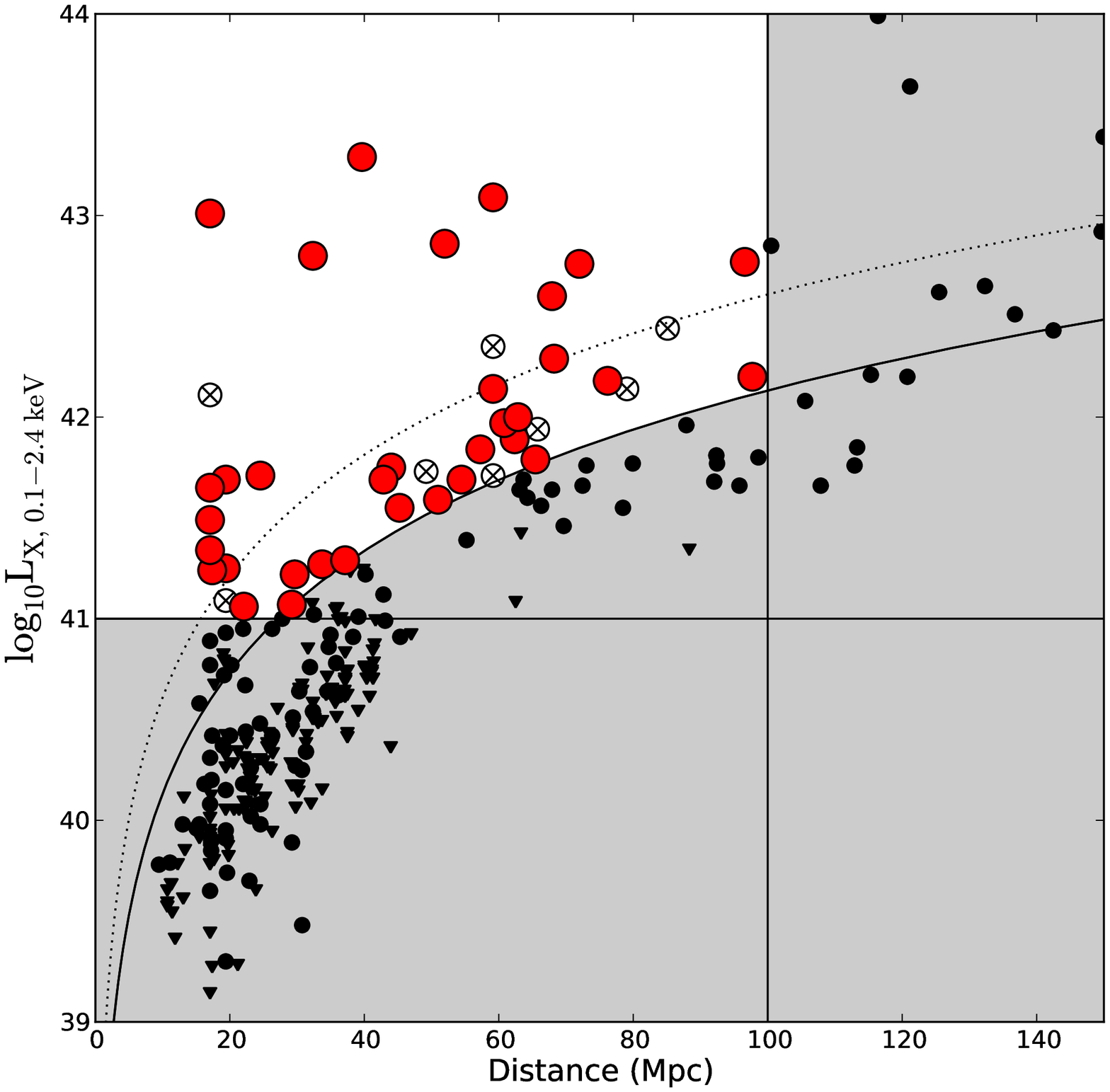}
\caption{\label{fig:extended} The distance-luminosity distribution of
  the galaxy sample of \citealt{Beuing99} with the lower flux cut and
  the luminosity cut used to create the extended sample.  Those with
  detected radio emission in the NVSS or SUMSS have 
a red circle, those with no detection have a crossed,empty circle.  We show
the distance and X-ray luminosity limits, as well as the old (dotted)
and new (solid) flux limits used to create this sample.} 
\end{figure}

Table \ref{tab:extend} summarises the radio information from the NVSS
and SUMSS for all galaxies in the extended sample.   It is clear that this
extended sample still has a very high radio detection rate, with 34/42
(81 per cent) galaxies having a radio detection within NVSS/SUMSS.
The radio sensitivity of this analysis is much lower than that in the
previous sections, using pointed VLA observations.  Therefore the
total fraction could be significantly higher.  In Fig. \ref{fig:extended} we show the distribution of galaxies
and those which have radio detections in the NVSS and SUMSS. 

\begin{table}
\caption{Extended Sample\label{tab:extend}}
\begin{tabular}{lcccrccc}
\hline
\hline
Galaxy & Flux & Survey & Type & Redshift & $L_{X}$\\
&mJy&&&&$\log_{10}~\ergps$\\
\hline
IC310$^*$&	 168.1 &  N	&  GiC	  &   0.0156 & 42.60 \\
IC1860$^*$&	 18.3  &  N	&  GiC	  &   0.0221 & 42.77 \\
IC2006 	&	 -     &  N	&  GiC	  &   0.0045 & $<$41.09$^a$ \\
IC4296 	&	 546.6 &  N	&  rG 	  &   0.0118 & 41.59 \\
IC4765  &	 32.4  &  S	&  GiG	  &   0.0144 & 41.89 \\
NGC57 	&	 -     &  N	&  G 	  &   0.0136 & 41.71 \\
NGC410  &	 5.8   &  N	&  GiG	  &   0.0140 & 41.97 \\
NGC499$^*$&	 -     &  N	&  GiG	  &   0.0136 & 42.35 \\
NGC507$^*$&	 61.7  &  N	&  IG 	  &   0.0166 & 42.76 \\
NGC533$^*$&	 28.6  &  N	&  GiG	  &   0.0157 & 42.29 \\
NGC708$^*$&	 65.7  &  N	&  rG 	  &   0.0136 & 43.09 \\
NGC741  &	 478.8 &  N	&  rG 	  &   0.0151 & 41.79 \\
NGC777  &	 7.0   &  N	&  Sy2	  &   0.0136 & 42.14 \\
NGC1399$^*$&	 208.0 &  N	&  GiC	  &   0.0045 & 41.69 \\
NGC1404$^*$&	 3.9   &  N	&  GiC	  &   0.0045 & 41.25 \\
NGC1407 &	 87.7  &  N	&  GiG	  &   0.0051 & 41.06 \\
NGC1550$^*$&	 16.6  &  N	&  G 	  &   0.0120 & 42.86 \\
NGC2300 &	 2.9   &  N	&  IG 	  &   0.0069 & 41.22 \\
NGC2305	&	 -     &  S	&  G 	  &   0.0114 & 41.73 \\
NGC2329 &	 363.7 &  N	&  rG 	  &   0.0175 & 42.18 \\
NGC2340	&	 -     &  N	&  G 	  &   0.0182 & 42.14 \\
NGC3091 &	 2.5	& N	&  GiG	  &   0.0126 & 41.69 \\
NGC4073	&	-	& N	&  GiG	  &   0.0195 & 42.44 \\
NGC4203$^*$&	 6.1	& N	&  LIN	  &   0.0040 & 41.24 \\
NGC4261	&	 4066.7 & N	&  LIN	  &   0.0078 & 41.27 \\
NGC4406$^*$&	-	& N	&  GiC	  &   0.0040 & 42.11 \\
NGC4472$^*$&	 219.9  & N	&  Sy2	  &   0.0040 & 41.49 \\
NGC4486$^*$&	138487.0& N	&  LIN	  &   0.0040 & 43.01 \\
NGC4636$^*$&	 77.8	& N	&  LIN	  &   0.0040 & 41.65 \\
NGC4649$^*$&	 29.1	& N	&  GiP	  &   0.0040 & 41.34 \\
NGC4696$^*$&	 5674.0 & S	&  GiC	  &   0.0092 & 43.29 \\
NGC4936 &	 39.8	& N	&  LIN	  &   0.0102 & 41.75 \\
NGC5044$^*$&	 34.7	& N	&  GiG	  &   0.0075 & 42.80 \\
NGC5090 &	 1325.8 & S	&  IG 	  &   0.0105 & 41.55 \\
NGC5129 &	 7.2	& N	&  GiG	  &   0.0224 & 42.20 \\
NGC5328	&	-	& N	&  GiC	  &   0.0152 & 41.94 \\
NGC5419 &	 349.2  & N	&  rG 	  &   0.0132 & 41.84 \\
NGC5846$^*$&	 21.0	& N	&  GiP	  &   0.0057 & 41.71 \\
NGC6407 &	 54.6	& S	&  G 	  &   0.0145 & 42.00 \\
NGC6868 &	 138.8  & S	&  GiG	  &   0.0086 & 41.29 \\
NGC7049 &	 93.2	& S	&  GiC	  &   0.0068 & 41.07 \\
NGC7619 &	 20.3	& N	&  GiG	  &   0.0099 & 41.69 \\
\hline
Total & 34 & 42\\
\hline
\end{tabular}
\begin{quote}$^*$ - those galaxies in the detailed sample.  The type-codes are as follows G-Galaxy,
  GiC-Galaxy in Cluster, GiG-Galaxy in Group, GiP-Galaxy in Pair,
  rG-Radio Galaxy, IG-Interacting Galaxy(ies), Sy2-Seyfert 2 galaxy, LIN-LINER-type Active
  Galaxy Nucleus.  The surveys used are N-NVSS and S-SUMSS.  $^a$ IC\,2006 only has an upper limit on the X-ray luminosity
  so whether it is truly part of this sample is debatable.  It also
  has no radio detection, and so it is possible that, given the high
  radio detection rate, that it has a much smaller X-ray luminosity
  than shown in the table.  
\end{quote}
\end{table}

For the extended sample we are not able to comment on the morphology of the
X-ray emission, and only very coarsely on the radio morphology.
However, we can use the classifications given to the galaxies in the
SIMBAD\footnote{\url{http://simweb.u-strasbg.fr/simbad/}} database,
which are noted in Table \ref{tab:extend}.  There are no galaxies with
classifications which would indicate that they have radio emission
that we have overlooked/missed with the two surveys used.  The
galaxies which do not have any radio emission are a mixture of field
galaxies, as well as those in groups or clusters, but not dominating
them.

We investigated whether from the SIMBAD morphological markers we could
clearly determine whether the X-ray emission is likely to be dominated
by an X-ray bright AGN.  However NGC\,4472, NGC\,4486 and NGC\,4636
have markers of either Seyfert 2 galaxies or a LINER type AGN but all
have clear and strong diffuse X-ray emission in addition to the AGN
emission.  Without high spatial resolution X-ray observations of these
galaxies we cannot comment on the ratio of the nuclear to the diffuse
X-ray emission in these galaxies.

\section{Luminosity Function}\label{sec:lumFxn}

\begin{figure}
\centering
\includegraphics[width=0.49\textwidth]{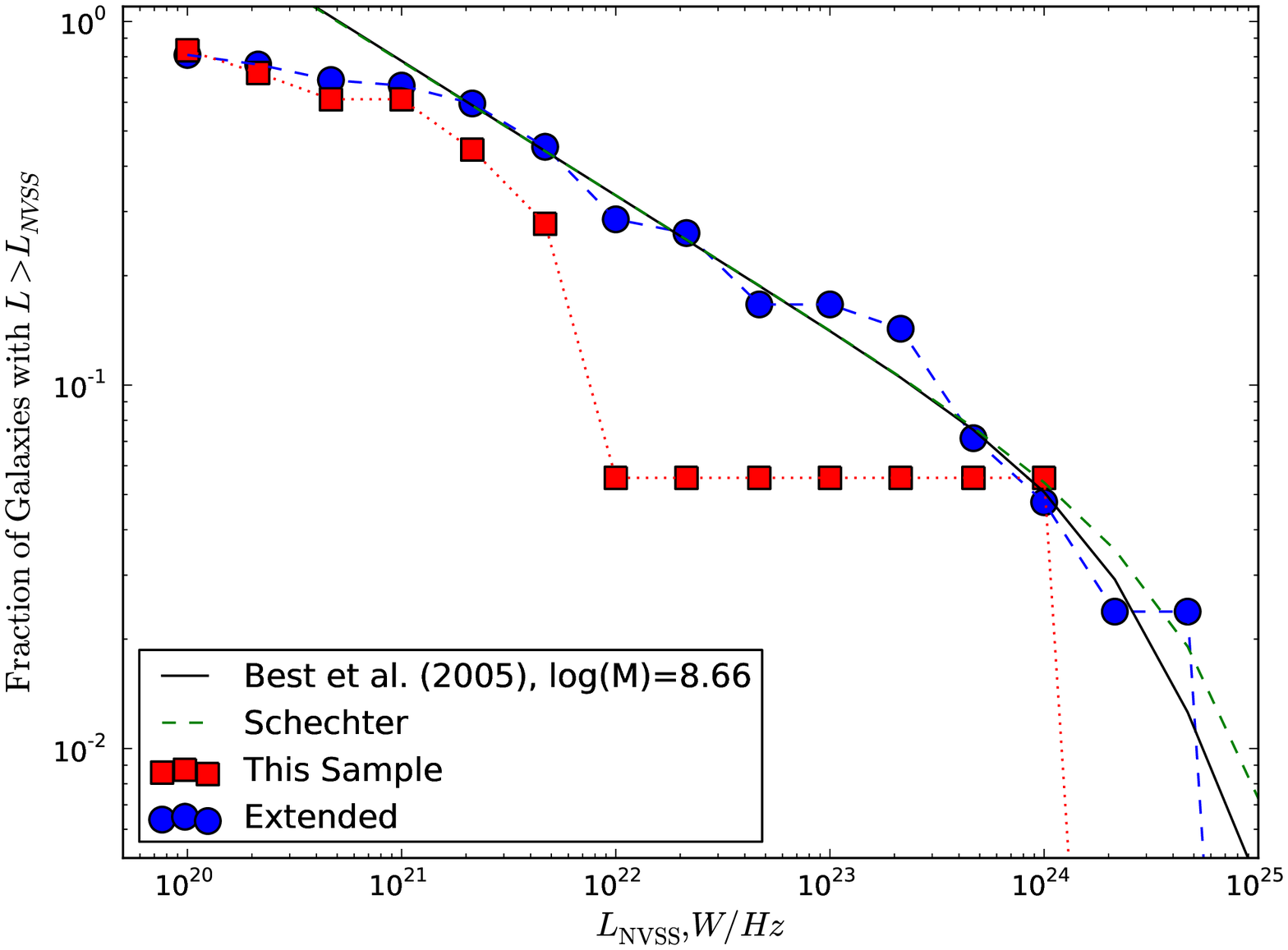}
\caption{\label{fig:LF} The luminosity functions of the main (red
  squares) and
  extended (blue circles) samples using the NVSS radio luminosities, supplemented by
  the SUMSS catalogue in the extended sample.  We also show the best
  fit relation to the extended sample using both a Schechter function (green line) and the form of
  the luminosity function shown in \citet{Best05}
  scaled for a mass of $10^{8.66} M_{\odot}$. as the solid black line.} 
\end{figure}

Fig. \ref{fig:LF} shows the cumulative radio luminosity functions for
the main and extended galaxy samples.  The data $>1\times 10^{21}$ have been fitted
with a simple Schechter function,
\begin{equation}
f_{\rm radio-loud}=f_0 \left( \frac{L}{L_*} \right)^\alpha \exp \left( \frac{L}{L_*} \right).
\end{equation}
A best fit results in $f_0=0.0289$, $\alpha=-0.367$ and
$L_*=7.79\times 10^{24}$ W/Hz for the bins $>1\times 10^{21}$.  
Below around $\sim 2\times 10^{21}$ W/Hz the function breaks again and
can be modelled with a powerlaw of slope $-0.1$.

We also use the model used by \citet{Best05}
\begin{equation}
f_{\rm radio-loud}=f_0 \left( \frac{M}{10^8 M_\odot} \right) \left[ \left(
  \frac{L}{L_*} \right) ^\beta + \left( \frac{L}{L_*} \right)
  ^\gamma \right] ^{-1},
\end{equation}
where $f_0=3.5\times 10^{-3}$, $\alpha=1.6$, $\beta=0.37$,
$\gamma=1.79$ and $L_*=3.2\times 10^{24}~{\rm W/Hz}$ but scaled for a
mass of $10^{8.66} M_{\odot}$.  In our study we have not 
split the sources by black hole mass, as in this case both samples are
too small for this to be appropriate.  Unsurprisingly the luminosity
function is not very smooth, especially for the main sample presented
in this paper.  Despite this, the behaviour of the luminosity function of
the radio galaxies shown in \citet{Best05},
with the break at around $10^{24} {\rm W/Hz}$, fits the luminosity
function well.  

When using both models, there appears to be evidence for a second break around
$2 \times 10^{21} {\rm W/Hz}$, and the behaviour below this break appears to
show that the fraction continues to rise towards one.  This second
break, if true, occurs at lower luminosities than those present in
\citet{Best05}, but they do see a flattening of the luminosity
function for the case of the most massive black hole bins.

Given the close agreement between the shapes of the luminosity
functions presented here and in \citet{Best05}, the galaxies in our
samples do not appear to be special -- at least in their radio
luminosities.  Furthermore, given the comparatively few galaxies in
our sample (42) compared to the many more in the SDSS study (2215) and
the good agreement between the two shapes of the radio luminosity
function, this implies that our sample is fairly unbiased to the radio
luminosity of the galaxies within it.

\section{Conclusions}\label{sec:concl}

Using a parent catalogue of elliptical galaxies with well determined
optical properties we have identified a sample of 18 nearby, X-ray bright
elliptical galaxies.  Out of these 18 galaxies, 17/18 (94 per cent)
have detected central radio emission and over half (10/18) have
extended radio structures larger than 1 arcsecond.  12
galaxies have clear evidence for some disturbance of their X-ray
halos, nine of which also have extended radio emission indicating a
likely interaction of the radio and X-ray plasmas.  Only two of
the galaxies in our sample are above the radio loudness criterion of
\citet{Best05} (2/18, 11 per cent). However, the true level of radio
activity is much higher. 
Extending the sample to lower X-ray fluxes, at the expense of data
quality and completeness we find that 34/42 galaxies have a radio
detection with the NVSS and SUMSS surveys.   The lower sensitivity of these two
surveys means this percentage should be regarded as a lower limit.

The mass-weighted cooling time within $1\kpc$ appears to anticorrelate
with the total radio luminosity, particularly once systems for which
the X-ray emission is dominated by nuclear point sources (IC\,310 and
NGC\,4203) are excluded.  This lends further support to the idea that
hot gaseous halos of elliptical galaxies provide an important fuel
source for the central AGN.

We have calculated upper limits on $k/f$, where $k$ is the ratio of the total
particle energy to that of relativistic electrons radiating in the
range of $10\mhz$ and $10\ghz$ and $f$ is the volume filling factor of
the plasma in the cavity, for the extended radio lobes in the sample.
The distribution of the upper limits on $k/f$ is broad and consistent with what would be
expected as a radio plasma ages. 

We calculate an X-ray radio loudness parameter for the total galactic
emission in these two bands, $\mathcal{GR}_{\rm X}$.  There is no
clear correlation of $\mathcal{GR}_{\rm X}$ with the galaxy
environment or $L_{\rm X}$.

\section*{Acknowledgements}

This research was supported
by the DFG cluster of excellence Origin and Structure of the
Universe (www.universe-cluster.de) and in
part by the U.S. Department of Energy under contract number
DE-AC02-76SF00515.  

RJHD thanks the Alexander von
Humboldt Foundation/Stiftung for financial support.  SWA acknowledges support for this work from the National Aeronautics
and Space Administration through Chandra Award Numbers AR7-8007X and
G09-0088X.  GBT acknowledges support for this work from the National Aeronautics
and Space Administration through Chandra Award Number GO7-8124X
issued by the Chandra X-ray Observatory Center, which is operated by
the Smithsonian Astrophysical Observatory on behalf of the National
Aeronautics and Space Administration under contract NAS803060A. GG is a
postdoctoral researcher of the FWO-Vlaanderen (Belgium). 

We thank
Frazer Owen for the radio map of NGC\,4486; 
Steve Rawlings and 
Paul Alexander for interesting discussions on $k/f$ and the anonymous
referee whose suggestions improved this paper. This research has
made use of the NASA/IPAC Extragalactic Data Base (NED) which is
operated by the JPL, California Institute of Technology, under
contract with the National Aeronautics and Space Administration.

\bibliographystyle{mn2e} 
\bibliography{mn-jour,bondi}

\end{document}